\documentclass[pra,10pt,twocolumn,superscriptaddress]{revtex4-1}
\usepackage{amsmath}
\usepackage{latexsym}
\usepackage{amssymb}
\usepackage{graphics,epstopdf}
\usepackage{hyperref}
\hypersetup{
    colorlinks = true,
    linkcolor =blue,
	citecolor=blue, 
	urlcolor=blue 
}

\newcommand{\ed}{\end{document}}
\newcommand{\beq}{\begin{equation}}
\newcommand{\eeq}{\end{equation}}

\usepackage{mathtools}
\DeclarePairedDelimiter{\ceil}{\lceil}{\rceil}

\begin{document}
\title{More current with less particles due to power-law hopping}
\author{Madhumita Saha}
\email{madhumita.saha91@gmail.com }
\affiliation{Physics and Applied Mathematics Unit, Indian Statistical
Institute, 203 Barrackpore Trunk Road, Kolkata-700 108, India}
\author{Archak Purkayastha}
\email{archakp2@gmail.com}
\affiliation{Department of Physics, Trinity College Dublin, Dublin 2, Ireland}
\author{Santanu K. Maiti}
\email{santanu.maiti@isical.ac.in}
\affiliation{Physics and Applied Mathematics Unit, Indian Statistical
Institute, 203 Barrackpore Trunk Road, Kolkata-700 108, India}

\begin{abstract}
We reveal interesting universal transport behavior of ordered one-dimensional fermionic systems with power-law hopping. We restrict ourselves to the case where the power-law decay exponent $\alpha>1$, so that the thermodynamic limit is well-defined. We explore the quantum phase-diagram of the non-interacting model in terms of the zero temperature Drude weight, which can be analytically calculated. Most interestingly, we reveal that for $1<\alpha<2$, there is a phase where the zero temperature Drude weight diverges as filling fraction goes to zero. Thus, in this regime, counter intuitively, reducing number of particles increases transport and is maximum for a sub-extensive number of particles. Being a statement about zero-filling, this transport behavior is immune to adding number conserving interaction terms. We have explicitly checked this using two different interacting systems. We propose that measurement of persistent current due to a flux through a mesoscopic ring with power-law hopping will give an experimental signature of this phase. In persistent current, the signature of this phase survives up to a finite temperature for a finite system. At higher temperatures, a crossover is seen.  The maximum persistent current shows a power-law decay at high temperatures. This is in contrast with short ranged systems, where the persistent current decays exponentially with temperature. 
\end{abstract}

\maketitle
\subsection{Introduction}

Quantum systems with long-range interactions present a platform with a lot of rich and interesting physics which differ greatly from the corresponding short-ranged systems. Recent controlled experimental realization of long-range quantum spin systems on several platforms such as Rydberg atoms \cite{expt_Rydberg_atoms1,expt_Rydberg_atoms2,expt_Rydberg_atoms3}, trapped ions \cite{expt_trapped_ions1,
expt_trapped_ions2,expt_trapped_ions3,expt_trapped_ions4,expt_trapped_ions5,
expt_trapped_ions6,expt_trapped_ions7,expt_time_crystal2,Experiment_transport}, polar molecules \cite{expt_polar_molecules1,expt_polar_molecules2,expt_polar_molecules1,
expt_polar_molecules3}, dipolar gas \cite{expt_dipolar_gas1}, nuclear spins \cite{expt_nuclear_spins}, nitrogen-vacancy centers in diamond \cite{expt_time_crystal1} and trapped atoms \cite{expt_cold_atoms_in_waveguides1,expt_atoms_in_trap}, and demonstration of  exotic physics such as time crystals \cite{expt_time_crystal1,expt_time_crystal2}, prethermalization\cite{expt_trapped_ions3}, dynamical phase transitions\cite{expt_nuclear_spins,expt_trapped_ions2,expt_trapped_ions7,expt_atoms_in_trap} and environment assisted transport \cite{Experiment_transport} have sparked intense research activities in such systems. Theoretical investigations are going on in the exotic physics mentioned above \cite{theory_DQPT1,theory_DQPT2,theory_pretherm1,theory_pretherm2,
theory_pretherm3,theory_power_law14,theory_power_law15,alessio_kaptiza_2018}, and also on exploring Lieb-Robinson bounds\cite{theory_LRB1,theory_LRB2,theory_LRB3,theory_LRB4,theory_LRB5,theory_LRB6,
theory_LRB7,theory_LRB8,theory_LRB9,Laurent}, entanglement dynamics\cite{Ranjan_Tanay,theory_entanglement1,
theory_entanglement2,theory_entanglement3,theory_entanglement4,theory_entanglement5,
theory_entanglement6,theory_entanglement7}, and localization properties\cite{theory_old_localization1,theory_old_localization2,
theory_old_localization3,theory_old_localization4,theory_old_localization5,
theory_old_localization6,theory_power_law_loc1,theory_power_law_loc2,theory_power_law_loc3,
theory_power_law_loc4,theory_power_law_loc5,theory_power_law_loc6,theory_power_law_loc7,
theory_power_law_loc8,theory_power_law_loc9,theory_power_law_loc10,
theory_power_law_loc12,theory_power_law_loc13,
theory_mbl_long_range0,theory_mbl_long_range1,theory_mbl_long_range2,
theory_mbl_long_range3, theory_long_range_mbl5,
theory_long_range_mbl6,arti_garg_2019,theory_fully_connected1,
theory_fully_connected2,theory_fully_connected3} of long-range systems.  

The localization properties of long-range disordered systems have been in limelight recently. In one and two dimensions, short-range non-interacting (quadratic Hamiltonian) systems with random on-site disorder show Anderson localization. It was shown in early works that in presence of long-range hopping that decays as a power-law, it is possible to have delocalized states in low dimensions\cite{theory_old_localization1,theory_old_localization2,
theory_old_localization3,theory_old_localization4,theory_old_localization5,
theory_old_localization6}. Until recently, it was commonly believed that long-range hopping in disordered systems in low-dimensions have a delocalizing effect. But very recent works have shown that long-range correlated hopping can also lead to localization\cite{theory_power_law_loc1,theory_power_law_loc3}. Further works have explored the effect of power-law hopping on quasi-periodic systems\cite{Ranjan_Tanay, theory_power_law_loc4,theory_power_law_loc5}. Another interesting direction has been on the fate of many-body localization in presence of long-range hopping. Several very recent works explore this direction\cite{theory_mbl_long_range0,theory_mbl_long_range1,theory_mbl_long_range2,
theory_mbl_long_range3,theory_long_range_mbl5,
theory_long_range_mbl6,arti_garg_2019,theory_fully_connected1,
theory_fully_connected2,alessio_kaptiza_2018,theory_fully_connected3}.

The common theme that all these works point to is that low-dimensional long-range systems have very interesting transport properties. However, surprisingly, there is an extreme lack of works directly exploring transport coefficients of long-range systems. In Ref.\cite{energy_transport_theory1}, the energy transport in an Anderson insulator in presence of power-law interactions was explored. In this work, we directly explore transport properties of a one-dimensional fermionic ordered system with power-law hopping. We confine ourselves to the case where power-law hopping exponent $\alpha>1$, so that the thermodynamic limit is well-defined. We characterize transport by calculating the zero temperature Drude weight. Our most interesting result is that, $1<\alpha<2$ is a phase where the Drude weight diverges as filling fraction goes to zero. Thus transport in such systems increase on reducing the number of particles and is maximum for a sub-extensive number of particles. This rather peculiar behavior stems from a non-analyticity of the energy dispersion relation of the non-interacting model. Being a statement about zero-filling, this is generically true for any ordered system in presence of any interaction term that conserves the number of particles. As an experimental signature of this phase, we propose the measurement of persistent current through a mesoscopic ring with power-law hopping. Persistent current has been previously measured in mesoscopic metal rings\cite{Persistent_experiment1,Persistent_experiment2,Persistent_experiment3,
Persistent_experiment4}. For mesoscopic ordered systems with power-law hopping, in the phase with $1<\alpha<2$, the persistent current will be greatly enhanced on reducing the number of particles, which is the tell-tale signature of this phase. The enhancement of persistent current on reducing the number of particles will be seen up to a finite-temperature for finite size systems. We also show that persistent current in such systems is exponentially larger than that in   short range systems. The maximum persistent current decays as a power-law with temperature, whereas in short-range systems, it is known to decay exponentially.
 
The paper is organized as follows. In Sec. B, the definitions of Drude weight and persistent current are introduced. The long-range hopping model and the existence of thermodynamic limit of that model are shown in Sec. C. Our main results concerning the zero temperature Drude weight are given in Sec.D. In Sec. E, we calculate persistent current for zero and finite temperature. In Sec. F, we conclude by summarizing our results and discussing further works.

\subsection{Definitions: The Drude weight and the persistent current}
The Drude weight characterizes the strength of the zero frequency peak in electrical conductivity. The electrical conductivity of an isolated system in the thermodynamic limit is given by the Green-Kubo formula. At finite frequency, the Green-Kubo formula is given by
\begin{align}
\sigma(\omega)=2\pi^2 D\delta(\omega)+\sigma^{reg}(\omega).
\end{align}
Here $D$ is the Drude weight, and $\sigma^{reg}(\omega)$ is the regular part of conductivity.  In a perfect conductor showing ballistic transport, $D$ is non-zero and conductivity diverges at zero frequency. Thus, for ballistic transport at zero frequency, the Drude weight is the only relevant quantity.  This is the case for non-interacting ordered systems, and for integrable systems in general.  

The Green-Kubo formula for conductivity is based on the linear response of the system to an electric field. However, there is another way of driving current in a system. Imagine a one-dimensional electronic system with periodic boundary conditions. Putting a magnetic flux $\Phi$ through the ring will drive a persistent current through the system. Let $F$ be the thermodynamic potential of the system. Then, the persistent current \cite{persistent1,persistent2,persistent3,persistent4,persistent5} through the system is given by
\begin{align}
\label{def_persistent}
& I(\Phi)=-\frac{\partial F}{\partial \Phi},  ~~ F=-T \log(Z),
\end{align}
where $T$ is the temperature. Here and henceforth we have set Boltzmann constant $k_B=1$.
At zero temperature $(T=0)$, the expression for persistent current reduces to 
\begin{align}
I(\Phi)=-\frac{\partial E_0}{\partial \Phi}
\end{align}
where $E_0$ is the ground state energy of the system.

The connection between these two different ways of driving transport was given by Kohn in his seminal work \cite{drude1}. He showed that, at zero temperature,
\begin{align}
\label{def_DN_D}
D(N)\equiv\frac{N}{4\pi^2}\frac{\partial^2 E_0}{\partial \Phi^2}{|_{\Phi=\Phi_{min}}},~~\lim_{N\rightarrow\infty} D(N) = D,
\end{align}
where, $\Phi_{min}$ is the flux at which the ground state energy $E_0$ is minimum and $N$ is the system size. For practical purposes, this gives an alternative, easier way to evaluate the zero temperature Drude weight for one dimensional systems. In the following, we will look at the Drude weight of a one-dimensional ordered systems with power-law hopping using this procedure. We will also look at the persistent currents in such systems. 

\subsection{The model and existence of the thermodynamic limit}

We consider a one-dimensional ordered non-interacting system with periodic boundary condition, with a flux $\Phi$ threaded through the ring. The flux $\Phi$ effectively modifies the hopping between sites at a distance $m$ by a Pierls' phase, $exp(i \frac{2 m \pi \Phi}{N \Phi_0})$. Here $\Phi_0$ is the flux quantum given by $\Phi_0$=$\frac{ c h}{e}$, where $c$, $h$, and $e$ are  velocity of light, Planck's constant and electronic charge. Henceforth, we will set $\Phi_0=1$. The resulting Hamiltonian is given by
\begin{align}
\hat{H} = -\sum\limits_{i=1}^N\sum\limits_{m=1}^{\ceil{N/2}-1} \left(\frac{ exp(i \frac{2\pi m \Phi}{N})}{m^{\alpha}} \hat{c}^{\dagger}_{i} \hat{c}_{i+m} + h.c\right)  
\label{hamil}
\end{align}
Here $\{c_i\}$ is the fermionic annihilation operator. The system has power-law hopping with exponent $\alpha$ and strength $-1$.   The long-range power-law hopping has a hard cut-off at $m=\ceil{N/2}-1$, where $\ceil{x}$ the least integer greater than or equal to $x$.  This cut-off is required to uniquely define the periodic boundary condition. The important point here is that the cut-off scales with system-size.

This model can be easily diagonalized via a Fourier transform. The single particle eigenvalues and eigenvectors of the system are given by
\begin{align}
& \varepsilon_{n} =-\sum_{m=1}^{\ceil{N/2}-1} \frac{2}{m^{\alpha}}\cos\left[\frac{2\pi m}{N}(n+\Phi)\right],  \nonumber \\
& \Psi_n(r)=\frac{1}{\sqrt{N}}exp(i\frac{2\pi n r}{N})           
\label{eigen}
\end{align}
where, $-N/2\leq n < N/2 -1$ when $N$ is even and $-\frac{(N-1)}{2}\leq n \leq\frac{(N-1)}{2}$ when $N$ is odd. Since the single particle eigenstates are all completely delocalized, transport is ballistic. So, as discussed above, transport is entirely characterized by the Drude weight.  The Drude weight is defined only in the thermodynamic limit. Since this is a long-range system, existence of the thermodynamic limit is not obvious.  It can be checked as follows that the thermodynamic limit is well-defined for $\alpha>1$. (In general, thermodynamic limit is well-defined for $\alpha>d$, where $d$ is the dimension of the system.)   

\begin{figure}
\includegraphics[width=\linewidth, height=4.8cm]{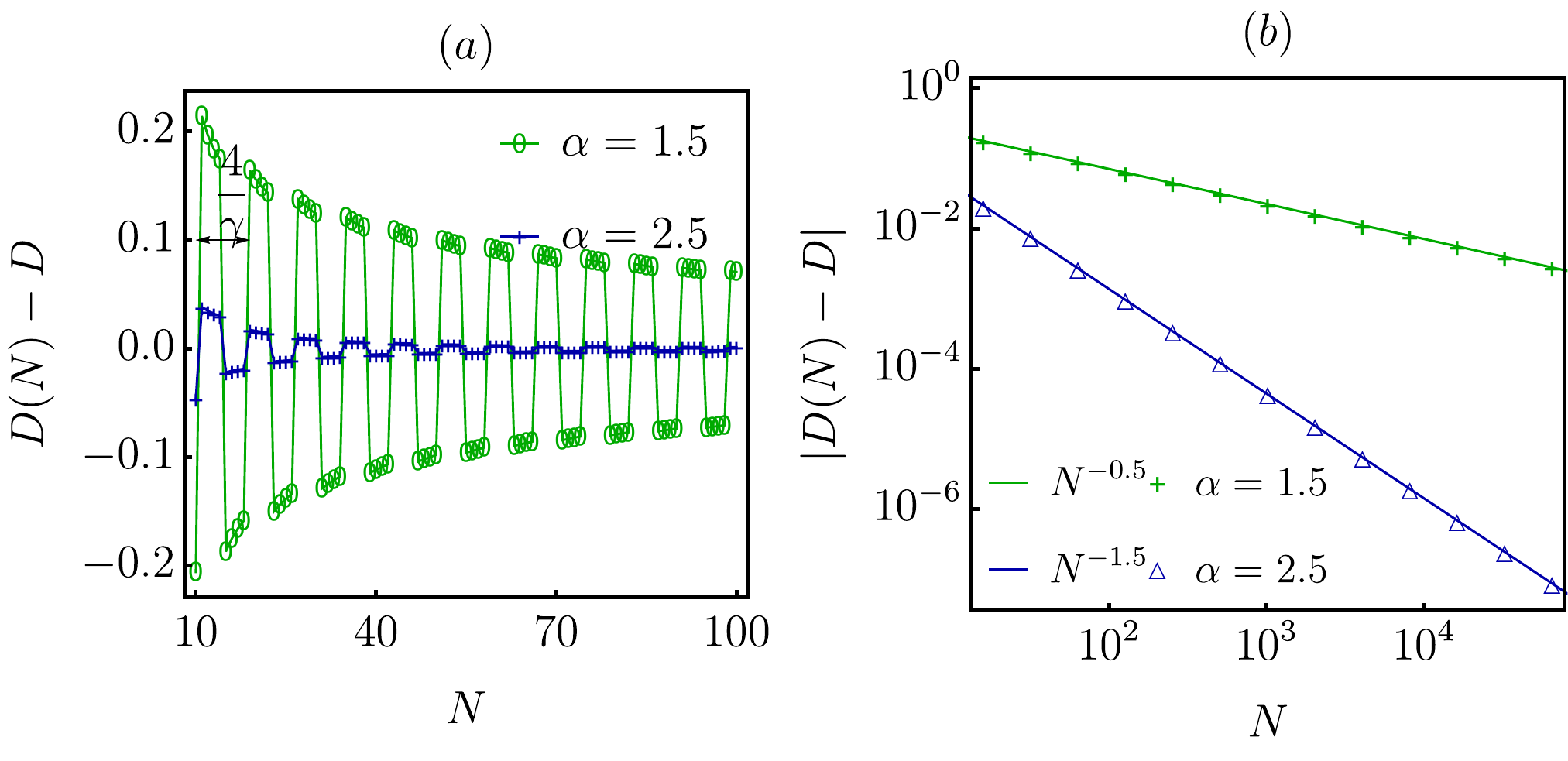}
\caption{(Color online). (a) The figure shows oscillations of $D(N)$ around $D$ with $N$  at half-filling. The period of oscillation is $4/\gamma=8$. The amplitude of oscillations decay with increase in system-size, as $D(N)$ approaches $D$. The continuous lines here are guide-to-eye.(b) The figure shows that the amplitude of oscillations decay as $N^{1-\alpha}$.} 
\label{fig1}
\end{figure}

For thermodynamic limit to exist, the summation in $\varepsilon_n$ must converge for $N\rightarrow \infty$. By standard convergence tests, this is guaranteed that the series $\sum_{m} \frac{1}{m^\alpha}$ converges. This is the hyperharmonic series, which converges for $\alpha>1$. Thus, the Hamiltonian is well-defined in the thermodynamic limit for $\alpha>1$. However this is not the only requirement for existence of the thermodynamic limit. At zero temperature, one has to check further that the ground state energy is extensive.  For a given number of particles $N_e$, the many-body ground state energy $E_0$ is given by summing over $\varepsilon_n$ up to the Fermi level. This can be done analytically to obtain,
\begin{align}
\label{E0}
E_0 =& -\sum_{m=1}^{\ceil{N/2}-1} \frac{2\cos({2m\pi \Phi^{\prime}}/{N})\sin({m\pi \gamma})}{m^\alpha\sin({m\pi}/{N})}, \nonumber \\
& -0.5\leq\Phi^{\prime}<0.5,~\gamma=\frac{N_e}{N}.
\end{align}
$\Phi^{\prime}=\Phi$ for odd $N_e$ and $\Phi^{\prime}=\Phi -0.5$ for even $N_e$, and $\gamma$ is the filling fraction. This relation always holds true for $\alpha>1$.  To check the extensivity of $E_0$, we note that the terms in the summation where $m\sim O(N)$ decay as $N^{-\alpha}$. So, for large $N$, the main contribution to the sum comes from small $m$. Thus, we can truncate the upper limit of the sum to $m=m^*$, where $m^*$ is independent of $N$. For such small $m$, we have $\sin({m\pi}/{N})\simeq {m\pi}/{N}$ So, upon taking the limit $N\rightarrow \infty, m^*\rightarrow \infty$, such that $m^*/N \ll 1$, we have
\begin{align}
\label{E0_extensive}
E_0 &\simeq -N \sum_{m=1}^{\infty} \frac{2}{\pi m^{\alpha+1}}\cos({2m\pi \Phi^{\prime}}/{N})\sin({m\pi \gamma}) \nonumber \\
& \simeq -N \sum_{m=1}^{\infty} \frac{2}{\pi m^{\alpha+1}}\sin({m\pi \gamma})\nonumber\\
& = -N \frac{2}{\pi} S_{\alpha+1}(\pi \gamma),
\end{align}
where
\begin{align}
\label{def_Clausen}
S_{z}(x)=\sum_{m=1}^{\infty} \frac{\sin(m x)}{m^z},
\end{align}
is the generalized Clausen function of order $z$.
Thus, for large $N$ and given filling $\gamma$, $E_0 \propto N$ and hence is indeed extensive for all values of $\alpha>1$. So, the thermodynamic limit is well-defined for $\alpha>1$. Since we are interested in calculating the Drude weight, we confine ourselves to $\alpha>1$.

\begin{figure}[t!]
\includegraphics[width=\linewidth, height=4.8cm]{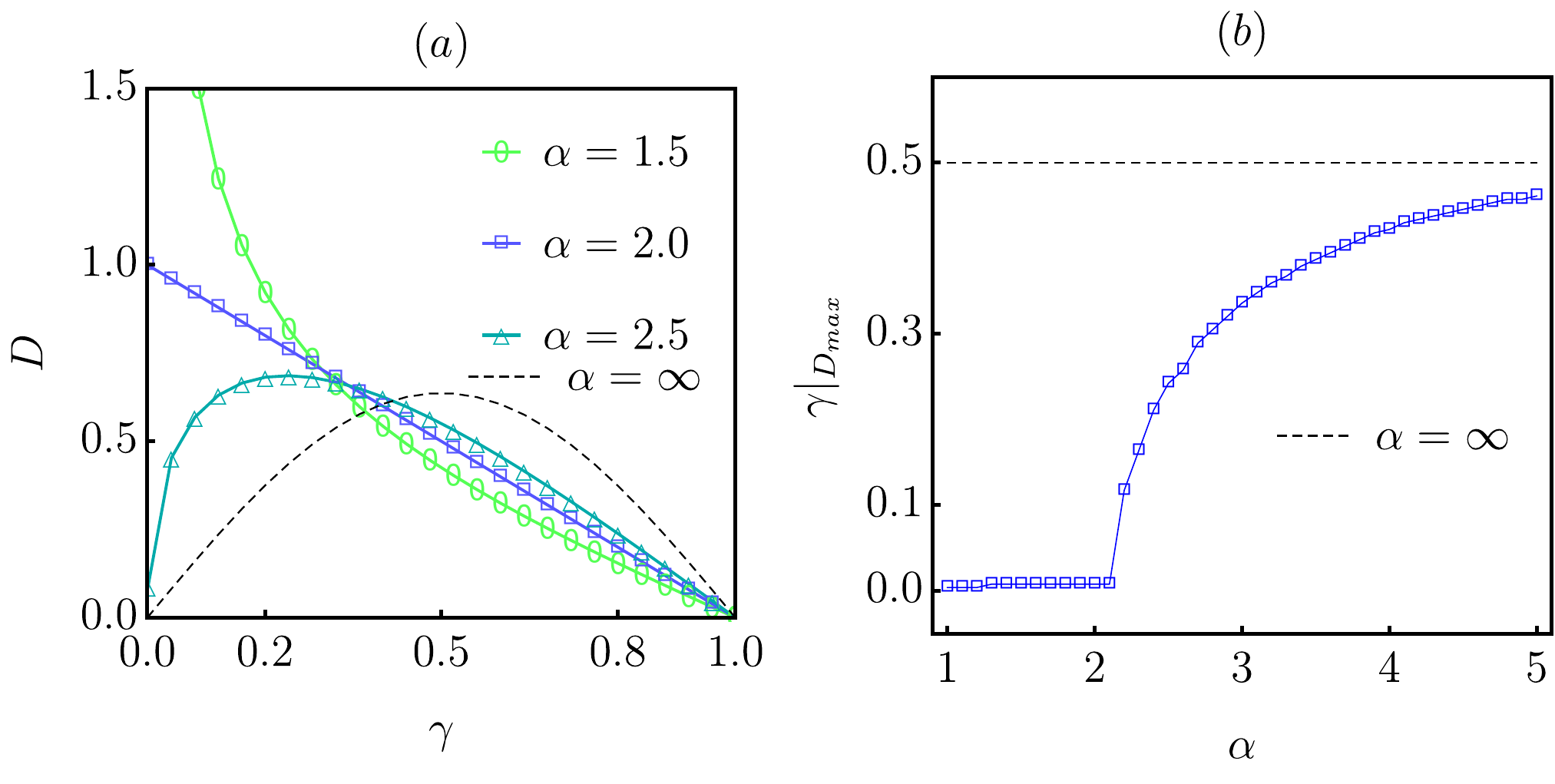}
\caption{(Color online) (a) The figure shows variation of the zero temperature Drude weight with the filling fraction for different values of $\alpha$. For $1<\alpha<2$, $D$ diverges as $\gamma\rightarrow 0^+$, and is discontinuous at $\gamma=0$. For $\alpha=2$, $D=1-\gamma$ for $\gamma>0$. Thus, for $1<\alpha\leq 2$, this the extremely counter-intuitive effect that transport increases on decreasing number of particles, and is maximum for a sub-extensive number of particles. For $\alpha>2$, $D$ has a maximum at a finite filling. (b) The filling at which $D$ is maximum is plotted with $\alpha$. The quantum phase transition at $\alpha=2$ is clear.  In both plots, $\alpha=\infty$ corresponds to the nearest neighbour hopping, where $D$ is maximum at half-filling. The continuous lines in both plots correspond to the analytical expression for $D$ (Eq.~\ref{D_analyt}), and the symbols correspond to $D(N)$ (Eq.~\ref{DN}) for $N=10^6$.   } 
\label{fig2}
\end{figure}

\subsection{The Drude weight: finite-size oscillations and divergence}
Armed with the expression for ground state energy, we now proceed to calculate the zero temperature Drude weight from Eq.~\ref{def_DN_D}. It can be checked that minimum value of $E_0$ occurs at $\Phi^\prime=0$.
Using Eq.~\ref{E0}, we can straightforwardly write down the expression for $D(N)$
\begin{align}
\label{DN}
D(N) = \frac{2}{N} \sum_{m=1}^{\ceil{N/2}-1} \frac{\sin({m\pi \gamma})}{m^{\alpha-2}\sin({m\pi}/{N})}.
\end{align}
The summation in the above expression can be broken into two parts: a) contribution from small $m$, (b) contribution from terms where $m\sim O(N)$. The contribution from small $m$ can be approximately calculated from the first line of Eq.~\ref{E0_extensive}. The contribution from terms with $m\sim O(N)$ is calculated by defining $x=m/N$. The resulting expression, for large $N$, is
\begin{align}
\label{DN_approx}
&D(N) \simeq \frac{2}{\pi} S_{\alpha-1}(\pi \gamma) + N^{1-\alpha} \sum_{x^*}^{1/2} \frac{\sin({x N\pi \gamma})}{x^{\alpha-2}\sin({x\pi})}, 
\end{align}
where $x^*<1/2$ is a finite number independent of $N$. The above expression exhibits that $D(N)$ shows oscillations with $N$. The period of the oscillation is given by that for $x=1/2$ in the summand in Eq.~\ref{DN_approx}. So the period of oscillation is $4/\gamma$. The amplitude of the oscillation decays as $N^{1-\alpha}$, for $\alpha>1$.  Thus, in the thermodynamic limit, we get the analytical expression for the Drude weight
\begin{align}
\label{D_analyt}
D=\frac{2}{\pi} S_{\alpha-1}(\pi \gamma).
\end{align}
Fig.~\ref{fig1}(a) shows the oscillations of $D(N)-D$ with $N$ for half-filling, for which the period of oscillation is $8$. Fig.~\ref{fig1}(b) shows the decay of $\mid D(N)-D\mid$ with $N$, showing the approach is $N^{1-\alpha}$. This establishes the correctness of the approximate expression in Eq.~\ref{DN_approx}.

So, we have found that the Drude weight is given in terms of the generalized Clausen function $S_z(x)$. Now, the following result can be shown for $S_z(x)$ (see Appendix~\ref{Sz_limit}),
\begin{align}
\label{Sz_result}
\lim_{x\rightarrow 0^+} S_z(x) \sim \lim_{x\rightarrow 0^+} x^{z-1}.
\end{align}
Thus, for $z<1$, $S_z(x)$ diverges at small $x$ as $x^{z-1}$. But, from the definition of $S_z(x)$, we see that $S_z(0)=0$. So, $S_z(x)$ is discontinuous at $x=0$ for $z<1$.  This means that for $1<\alpha<2$, the Drude weight diverges as $\gamma\rightarrow 0^+$ and it is discontinuous at $\gamma=0$. In fact, as shown in Fig.~\ref{fig2}(a) for $\alpha=1.5$, in this regime, $D$ decreases monotonically with $\gamma$ for $\gamma>0$.  This reveals the extremely counter-intuitive result that $1<\alpha<2$ describes a phase where transport increases with decrease in the number of particles, and is maximum for a sub-extensive number of particles. For $z>1$, $S_z(x)$ goes to zero at small $x$, and hence there is no discontinuity at $x=0$. As shown in Fig.~\ref{fig2}(a) for $\alpha=2.5$, for $\alpha>2$, $D$ shows the more intuitive behavior of having a maximum at a finite value of $\gamma$, i.e, for an extensive number of particles. {\it Thus, at $\alpha=2$, we get a quantum phase transition, with the Drude weight at zero temperature showing markedly different behavior on either sides of it}. This makes $\alpha=2$ the `critical' point of the phase transition. At $\alpha=2$, $D$ depends on $S_1(x)$. $S_1(x)$ is known in closed form,  $S_1(x)= (\pi-x)/2,~\forall~0< x < 2\pi$.  So we get the following simple expression for the Drude weight for $\alpha=2$,
\begin{align}
\label{D_alpha_2}
D=
\begin{cases}
& 1-\gamma,~\forall~\gamma>0,~~~\alpha=2 \\
& 0,~ \gamma=0.
\end{cases} 
\end{align}
Thus in the `critical' case, the discontinuity at $\gamma=0$ and the monotonic decrease with $\gamma$ survives, but, $D\rightarrow 1$ as $\gamma\rightarrow 0^+$. So even for this `critical' point, transport increases with decrease in the number of particles. The corresponding plot is shown in Fig.~\ref{fig2}(a). In Fig.~\ref{fig2}(b), the value of $\gamma$ where $D$ is maximum, is plotted against $\alpha$. The quantum phase transition at $\alpha=2$ is extremely clear from this plot.  The results for the nearest neighbour case ($\alpha=\infty$) are also shown in the plots for comparison. For nearest neighbour case, $D$ increases monotonically with increase in number of particles, and is maximum at half-filling due to particle-hole symmetry (which is not present in the long-range model).  
These observations can also be directly checked from Eq.~\ref{DN} for $D(N)$ (Appendix~\ref{DN_limit}).

The origin of the divergence of $D$ at $\gamma\rightarrow 0^+$ for $1<\alpha<2$ can be traced to a divergence of group velocity in the system in this regime.   In the thermodynamic limit, defining $k=\frac{2\pi n}{N}$, at zero flux, the dispersion formula can be written from Eq.~\ref{eigen} as,
\begin{align}
\varepsilon(k)=-\sum_{m=1}^{\infty} \frac{2}{m^{\alpha}}\cos(mk)
\end{align}
The group velocity $v(k)$ is given by the first derivative of $\varepsilon(k)$,
\begin{align}
v(k)=\frac{\partial \varepsilon}{\partial k}=\sum_{m=1}^{\infty} \frac{2}{m^{\alpha-1}}\sin (mk)=2 S_{\alpha-1}(k).
\end{align}
Thus, the behavior of $v(k)$ with $k$ is same as that of $D$ with $\gamma$.
This immediately shows that 
\begin{align}
\lim_{k\rightarrow 0} v(k)\rightarrow \infty~\forall~1<\alpha<2
\end{align}
Since $k=0$ is the bottom of the band, it corresponds to filling fraction going to zero. So, for $1<\alpha<2$, as the number of particle decreases, the group velocity of the particles increases, leading to a maximum transport for a sub-extensive number of particles. For $\alpha>2$, $\lim_{k\rightarrow 0} v(k)=0$, and this effect is not seen. 

An intuitive picture of why increasing particles in a system with long-range hopping can reduce transport, can be given as follows. Transport depends on the number of particles as well as the velocity of the particles. With a few number of particles, there are a lot of empty sites in the system. If the particles can hop to really long distances, this can lead to very high velocity, but of few particles. However, the fermionic nature of the particles inhibits the long range hopping as the number of particles increases. So, there is a trade-off between the number of particles and the range of hopping which dictates the filling at which there will be maximum transport. For $1<\alpha<2$, we find that the range of hopping is such that the velocity actually diverges for few particles. So increasing the number of particles only reduces the velocity, thereby reducing transport.

\begin{figure}[t!]
\includegraphics[height=4.4cm,width=8.8cm]{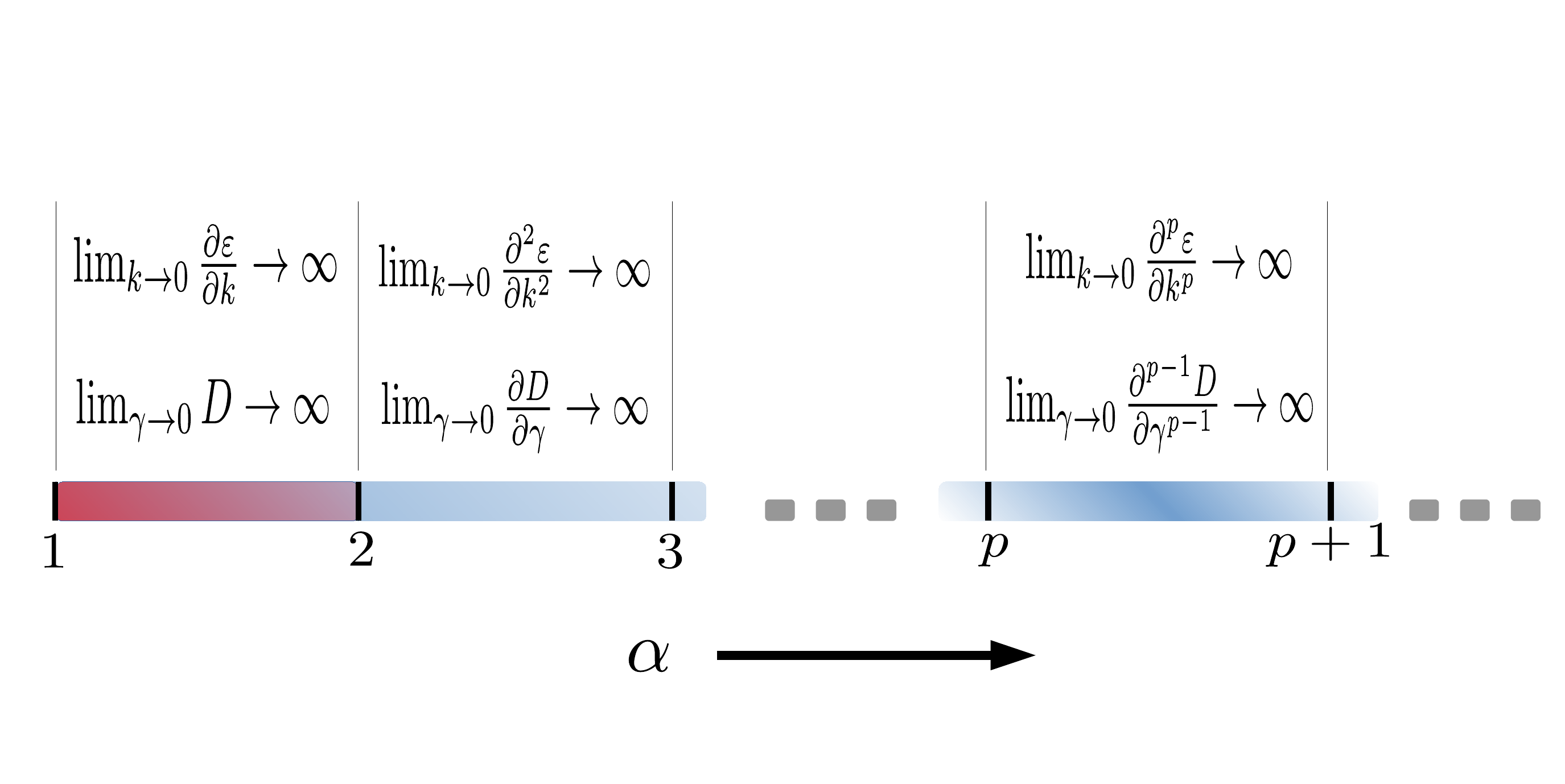}
\caption{(Color online) The dispersion relation for an ordered 1D non-interacting system with power-law hopping is non-analytic at $k=0$ for all values of power-law decay exponent $\alpha$ (see Eq.~\ref{dispersion_non_analytic}). The corresponding non-analyticity is reflected on the Drude weight $D$ at filling fraction $\gamma\rightarrow 0$ (see Eq.~\ref{drude_wieght_non_analytic}). The most spectacular is $1<\alpha<2$, where the $D$ itself diverges as $\gamma\rightarrow 0$. This behavior of Drude weight is completely immune to adding any interactions.} 
\label{fig:phase_diag}
\end{figure}

It is interesting to note that the dispersion relation $\varepsilon(k)$ for this system is actually non-analytic at $k=0$ for all values of $\alpha$. This can be seen as follows.
Let $p$ be an arbitrary integer. Then, it can be easily checked that, 
\begin{align}
\label{dispersion_non_analytic}
\lim_{k\rightarrow 0}\frac{\partial^p \varepsilon}{\partial k^p} \rightarrow \infty, ~\forall~\alpha<p+1.
\end{align}
This shows the non-analyticity of $\varepsilon(k)$ at $k=0$ for all values of $\alpha$. Thus, actually, for any integer $p\geq 1$, $p<\alpha<p+1$ is a phase marked by divergence of all derivatives of $\varepsilon(k)$ of higher order than $p$, and the integers $p$ and $p+1$ are the phase boundaries. This behavior can be seen in the Drude weight at filling fraction going to zero,
\begin{align}
\label{drude_wieght_non_analytic}
\lim_{\gamma\rightarrow0^+}\frac{\partial^{p-1} D}{\partial\gamma^{p-1}} \rightarrow \infty~\forall~\alpha<p+1
\end{align}
 However, the divergence of higher derivatives of $D$ does not affect the transport behavior as spectacularly as divergence of $D$ itself, which happens for $1<\alpha<2$ (see Fig.~\ref{fig:phase_diag}).

Now let us consider the effect of interactions. We consider only the type of interaction terms which conserve the number of particles. In other cases, it is no longer meaningful to consider systems at a given filling fraction.
In the limit of filling fraction going to zero, number-conserving interactions, however strong, play negligible role. To understand the last fact, consider the extreme case of the system with a single particle. In that case, there are no other particles in the system which it can interact with. So, the non-interacting system results are valid. In the thermodynamic limit, this case corresponds to zero filling fraction $\gamma \rightarrow 0^+$. Note, now, that in the thermodynamic limit, having any sub-extensive number of particles will also correspond to $\gamma \rightarrow 0^+$. So, as long as the thermodynamic limit exists, having any sub-extensive number of particles should be same as having a single particle, which, in turn is same as the non-interacting case with a single particle. For our case, this means that, \emph{the divergence of Drude weight for $1<\alpha<2$ at $\gamma\rightarrow 0^+$ is immune to presence of arbitrary interaction terms which conserve the number of particles.} 

\begin{figure}[t!]
\includegraphics[width=\columnwidth]{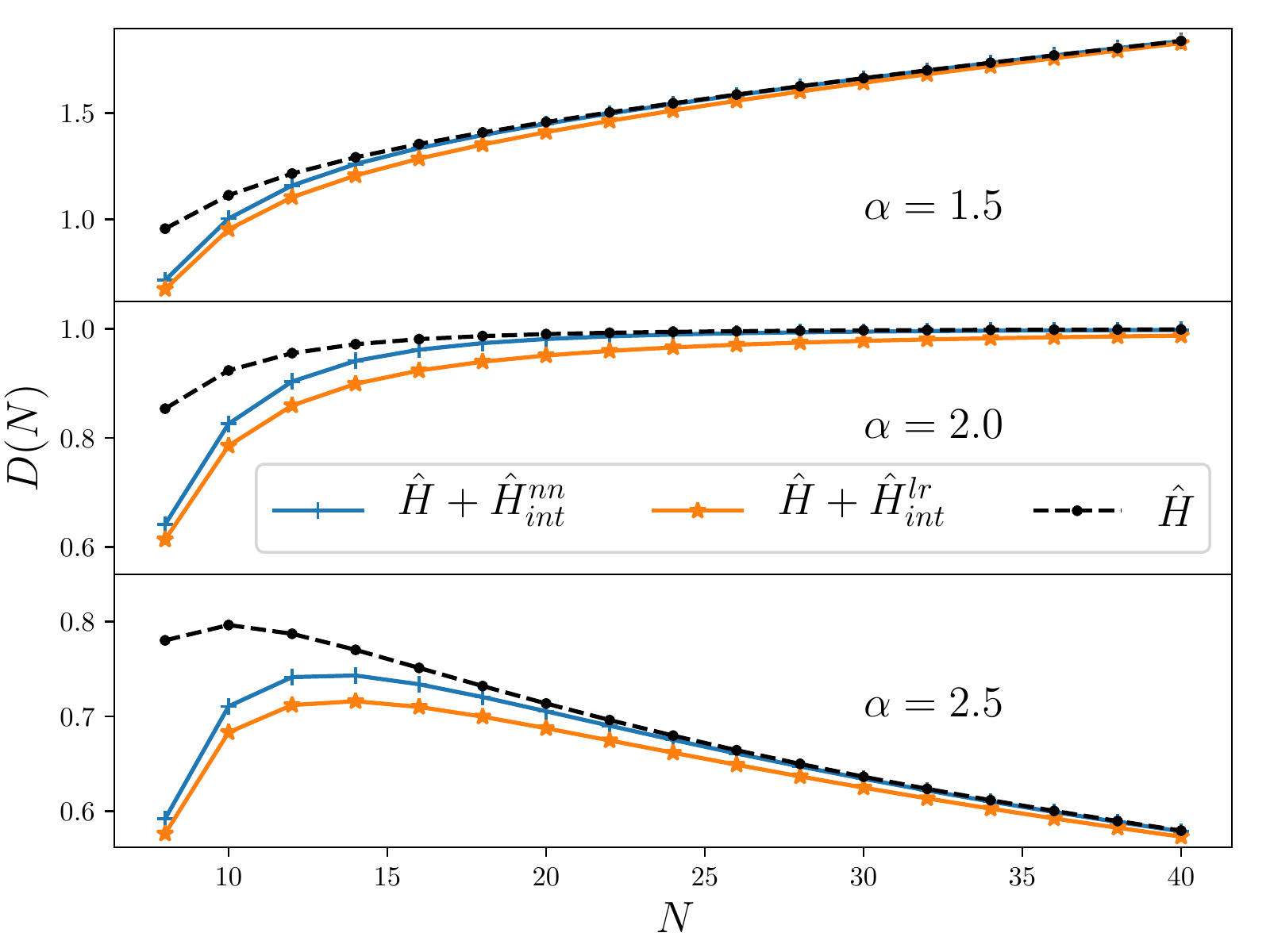}
\caption{(Color online)  The figure shows $D(N)$ vs $N$ in presence of nearest neighbour ($\hat{H}_{int}^{nn}$) and long-range interactions ($\hat{H}_{int}^{lr}$) (see Eq.~\ref{def_H_int}), in the 3 particle subspace ($N_e=3$), with strong interaction strengths ($V=10$), for three different values of $\alpha$: $\alpha=1.5$ (top), $\alpha=2.0$ (middle), $\alpha=2.5$ (bottom). Also shown are corresponding plots for the non-interacting long-range system $\hat{H}$. Since number of particles is sub-extensive, as expected, with increase in $N$, effect of interaction is lost. So, for $\alpha<2.0$, $D(N)$ diverges, for $\alpha=2.0$, $D(N)$ saturates to $1$, and for $\alpha>2.0$, $D(N)$ decays to zero.   } 
\label{fig:interacting_3_particles}
\end{figure}

We now check this explicitly taking two extremely different interacting systems. The two different models of interaction we consider are
\begin{align}
\label{def_H_int}
& \hat{H}_{int}^{nn} = V\sum_{\ell=1}^{N-1} \hat{n}_{\ell}\hat{n}_{\ell+1}    \nonumber \\
& \hat{H}_{int}^{lr} = V\sum_{i=1}^N\sum_{m=1}^{\ceil{N/2}-1} \frac{1}{m^{\alpha}} \hat{n}_{i} \hat{n}_{i+m},   
\end{align}  
where $\hat{n}_{i}=\hat{c}_i^\dagger \hat{c}_i $. The first of the two above interaction Hamiltonians describes nearest neighbour interactions, while the second describes long-range interactions with power-law decay. For simplicity, we have chosen the power-law exponent to be same as that in the non-interacting part. At a finite filling, i.e, for an extensive number of particles, these two models have extremely different properties (see, for example, \cite{theory_long_range_mbl6}). Here, we show that the divergence of Drude weight for a sub-extensive number of particles occurs in both models. So we fix the number of particles at $N_e=3$ and plot $D(N)$ vs $N$ in presence of above interactions, and compare with the non-interacting model. The plots
are shown in Fig.~\ref{fig:interacting_3_particles}. We have kept interactions strong, $V=10$. As expected, with increase in $N$ the effect of interactions is lost, and $D(N)$ for both the interacting models approach that of the non-interacting one. Thus, it is also clear from the plots, that even in presence of interactions, for $\alpha<2$, $D(N)$ diverges, for $\alpha=2.0$, $D(N)$ saturates to $1$, and for $\alpha>2.0$, $D(N)$ decays to zero, in the limit of zero filling fraction.   

The above reasoning, followed by the example of two extremely different interacting systems, strongly corroborates the fact that \emph{the divergence of zero temperature Drude weight for a sub-extensive number of particles is a general effect in ordered systems with power-law hopping with power-law exponent $1<\alpha<2$ in presence of arbitrary number conserving interactions.}   Having shown this, in the following we return to the non-interacting system and address the more pertinent question of experimental significance.

Although, increase in $D$ implies enhancement of transport, the experimental implication of this is not immediately clear in terms of electronic conductivity. This is because, any non-zero value of $D$ gives infinite electronic conductivity. So, in terms of electronic conductivity, the meaning of a divergence of $D$ is not obvious. A more direct experimental implication can be given in terms of the persistent current, which we explore in the next section.

\subsection{The persistent current}

\subsubsection{Zero temperature}

Directly from the definitions of the $I(\Phi)$ and $D(N)$, it is clear that system size behavior of $NI(\Phi)$ is same as that of $D(N)$. Since, for perfect conductors, $D(N)$ goes to a constant with system-size, $I(\Phi)$ decays as $N^{-1}$. (This is one of the ways of distinguishing a perfect conductor from a superconductor, where there can be persistent currents up to macroscopic length scales.) So observation of persistent current requires mesoscopic systems. The relation between  $NI(\Phi)$ at zero temperature and $D(N)$ precisely means that the physics described in previous sections can be experimentally explored by measuring persistent currents in mesoscopic rings with long-range hopping.

\begin{figure}
\includegraphics[width=\columnwidth]{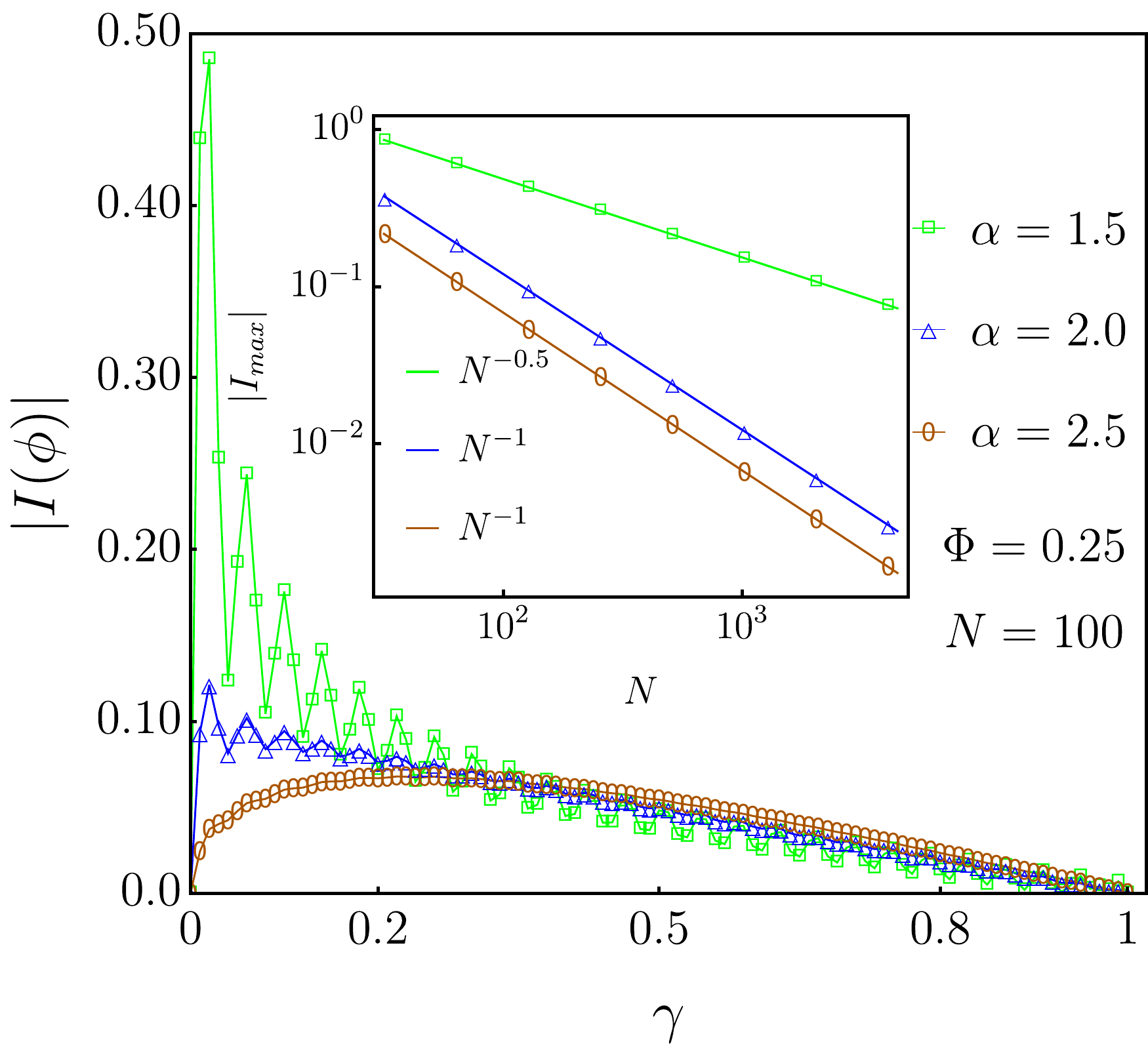}
\caption{(Color online) The figure shows behavior of magnitude of persistent current at zero temperature with the filling fraction for three different values of $\alpha$. We see that persistent current decreases with increase in number of particles for
 $1<\alpha\leq 2$. The enhancement is particularly more for $\alpha<2$. For $\alpha>2$, maximum persistent current occurs at a finite filling.  This is as expected from our investigation of Drude weight. The oscillations with $\gamma$ has a period of $4/N$, and are suppressed as $N^{-\alpha}$ with increasing $\alpha$. Here $\Phi=\Phi^{\prime}+0.5$ (See Eq.~\ref{I}). 
In inset the figure shows the absolute value of maximum current scaling with system size $N$ for three different $\alpha$. For 
$1<\alpha< 2$, it scales as $N^{1-\alpha}$ where the maximum current is resulting for sub-extensive number of particles. For $\alpha\geq 2$ it scales as $N^{-1}$.}
\label{fig3}
\end{figure}

To see this explicitly, let us write down the expression for persistent current at zero temperature from Eqs.~\ref{def_persistent} and \ref{E0},
\begin{align}
\label{I}
& I(\Phi^{\prime}) = -\frac{4\pi}{N} \sum_{m=1}^{\ceil{N/2}-1} \frac{\sin({2 m \pi \Phi^{\prime}/N})\sin({m\pi \gamma})}{m^{\alpha-1}\sin({m\pi}/{N})}, \\
& -0.5\leq\Phi^{\prime}<0.5, \nonumber
\end{align}
$\Phi^{\prime}=\Phi$ for odd number of particles and $\Phi^{\prime}=\Phi-0.5$ for even number of particles. Exactly as we have done for the Drude weight, the summation in the above equation can be broken in two parts corresponding to small $m$ and $m\sim O(N)$. For large $N$, the expression becomes
\begin{align}
\label{I_approx}
I(\Phi^{\prime}) \simeq \frac{8\pi\Phi^{\prime}}{N} S_{\alpha-1}(\pi \gamma) + \frac{N^{1-\alpha}}{N} \sum_{x^*}^{1/2} \frac{\sin({2 x\pi\Phi^{\prime}})\sin({x N\pi \gamma})}{x^{\alpha-1}\sin({x\pi})}
\end{align}
It is evident that behavior of $NI(\Phi^{\prime})$ with system-size is same as that of $D(N)$. We see from above expression that at finite filling $I(\Phi^{\prime})\sim N^{-1}$, as expected for ballistic system.  Also as expected from relation with $D(N)$, there are oscillations with system size with a period of $4/\gamma$, the amplitude of which decay as $N^{-\alpha}$. On the other hand, noting that $\gamma\sim 1/N$ for a sub-extensive number of particles, and using Eq.~\ref{Sz_result}, we see that
\begin{align}
\lim_{\gamma\rightarrow 0^+} I(\Phi^{\prime})\sim N^{1-\alpha},~\forall~\alpha>1
\end{align} 
This shows that $I(\Phi^{\prime})$ goes to zero in the thermodynamic limit even as $\gamma\rightarrow 0^+$, for all values of $\alpha>1$, which is the expected behavior for non-superconducting systems. Further, remembering that $I(\Phi^{\prime})\sim N^{-1}$ at finite filling, we see that, with $N$ fixed, for $1<\alpha<2$,  $I(\Phi^{\prime})$ is enhanced on reducing the number of particles. This, as discussed above, is the hallmark of this phase. This is true for any value of  $\Phi^{\prime}$ where $I(\Phi^{\prime})$ is non-zero. This also shows that for $1<\alpha<2$, the maximum persistent current decays exponentially with $\alpha$. Exactly at $\alpha=2$, the persistent current $\sim N^{-1}$ at all filling. But, from Eq.~\ref{I_approx}, it is seen that the maximum persistent current still occurs at $\gamma\rightarrow 0$. The variation of the magnitude of persistent current with $\gamma$ is shown for three values of $\alpha$ in Fig.~\ref{fig3} for $N=100$, at a chosen value of the flux. In the regime $1<\alpha\leq 2$, the enhancement of persistent current for a sub-extensive number of particles for is very clear.  The oscillations with $\gamma$ in Fig.~\ref{fig3} have period $4/N$, as can be seen from Eq.~\ref{I_approx}. These oscillations are suppressed with system size as $N^{-\alpha}$. The behavior of the maximum current $I_{max}$ is shown with system size in the inset for various values of $\alpha$. As explained above, it decays as $N^{1-\alpha}$ for $1<\alpha< 2$, and has $N^{-1}$ for $\alpha\geq 2$.

\begin{figure}
\includegraphics[width=8.5cm, height=4.0cm]{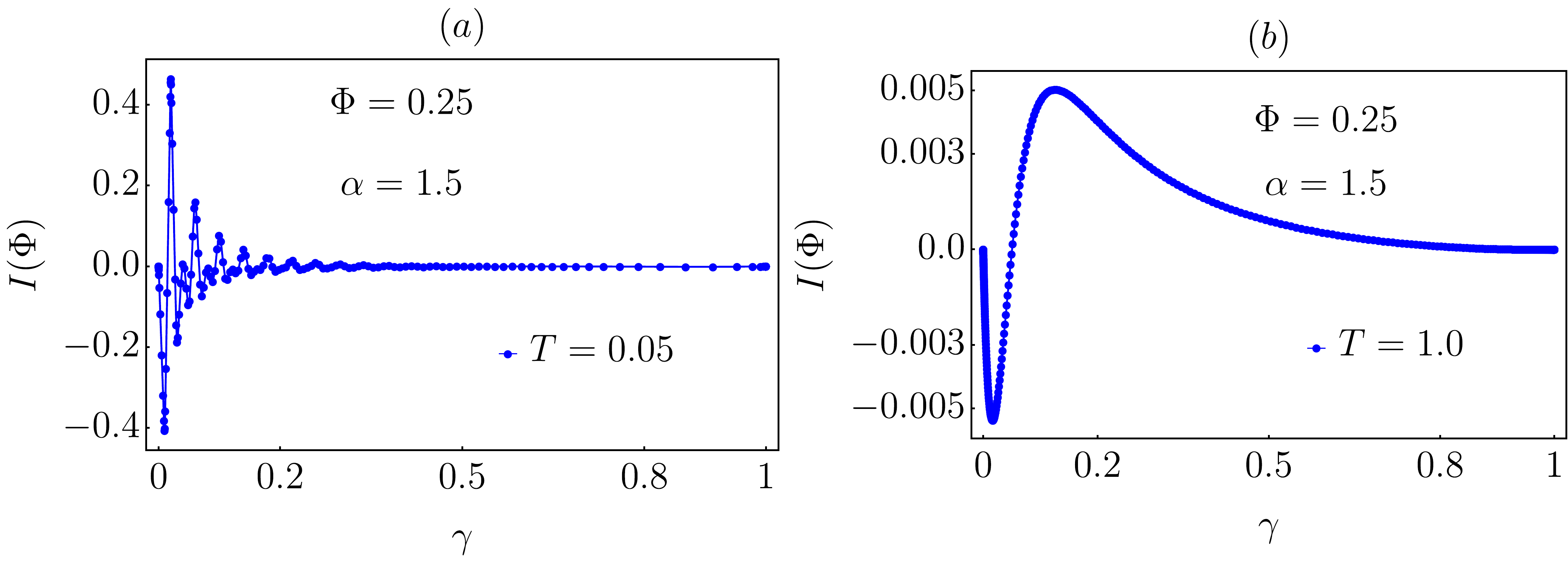}
\caption{(Color online) The figure shows variation of persistent current with filling fraction for a system of size $N=100$ and $\alpha$ at finite temperature (a) $T=0.05$, (b) $T=1.0$, for a chosen value of flux. At finite but low temperature, the maximum persistent current occurs at very small filling. But at higher temperatures, the maximum shifts to a finite filling. Also, from the y-scale of the plots, it is clear that persistent current decreases with increase in temperature. } 
\label{fig4}
\end{figure}

\begin{figure*}
\includegraphics[width=\linewidth]{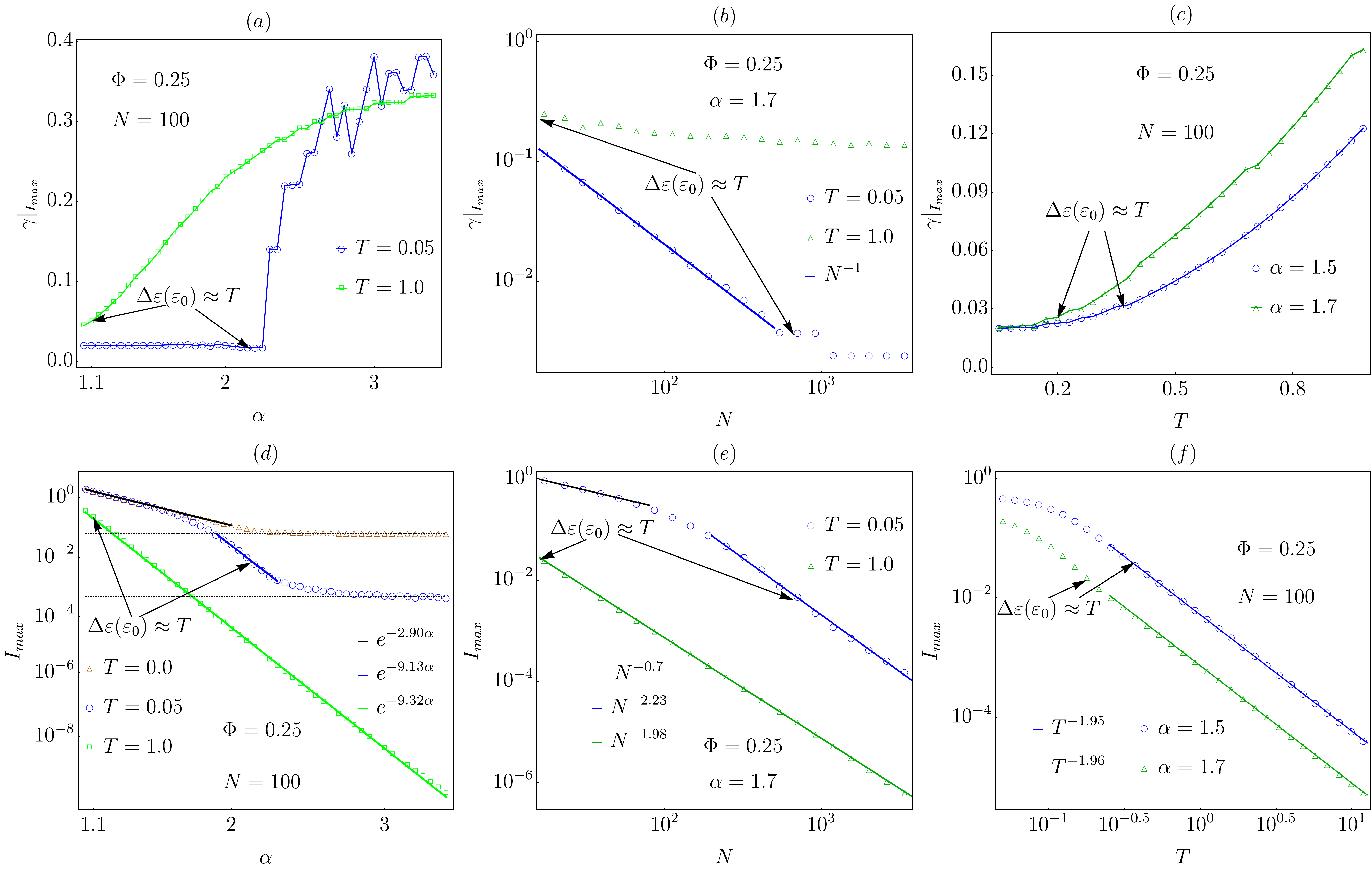}
\caption{(Color online) Top panel: Variation of $\gamma|_{I_{max}}$, the filling at which the maximum current occurs, with various parameters (a) with $\alpha$ at finite temperatures, (b) with system-size $N$ at finite temperatures, (c) with temperature $T$. Bottom panel: the corresponding variation in the magnitude of maximum persistent current $I_{max}$. Here $\Delta\varepsilon(\varepsilon_0)$ is the energy difference between the lowest single particle level and the next one. From (a),(b),(d) and (e) it is clear that the transition is seen in persistent current behavior  for $T\ll \Delta\varepsilon(\varepsilon_0)$. But as shown in (c) and (f), as a function of temperature there is no sharp transition. In (f) we see that the maximum persistent current decays as a power-law with temperature. This is in stark contrast with the nearest neighbour system, where it is known that persistent current decays exponentially with temperature.} 
\label{fig5}
\end{figure*}

So far we have shown that the phase showing the counter-intuitive feature of maximum transport at a sub-extensive number of particles is well captured by persistent current at zero temperature. In the next subsection, we check the effect of temperature.

\subsubsection{Finite temperature}
The effect of having maximum current at a sub-extensive number of particles is essentially due to large level-spacings near the bottom of the band for $1<\alpha\leq 2$. This means, for a finite system, the effect will survive as long as the temperature is small enough to resolve the energy level spacing at the bottom of the band. Thus, phase showing `more current with less particles' can be experimentally observed in the temperature regime
\begin{align}
\label{finite_temp_conditon}
T\ll \Delta\varepsilon(\varepsilon_0),
\end{align}
where $\Delta\varepsilon(\varepsilon_0)$ is the energy difference between the lowest single-particle level and the next one. It can be shown that, $\Delta\varepsilon(\varepsilon_0) \sim N^{1-\alpha}$. So, for a finite system, the effect can be observed from persistent current measurements up to a finite temperature. At higher temperatures, the level-spacing cannot be resolved, and the effect will be lost. This means, in the thermodynamic limit, there is no phase transition as a function of $\alpha$ at any finite temperature. This is typical for quantum phase transitions. Let us now explicitly check this.

We assume a grand-canonical ensemble with inverse temperature $\beta$ and chemical potential $\mu$. The persistent current and the filling fraction at a finite temperature and chemical potential is given by,
\begin{align}
\label{I_T}
& I(\Phi) = -\sum_{n} \frac{\partial\varepsilon_{n}}{\partial \Phi}f_{n},~ \gamma = \frac{1}{N}\sum_{n} f_{n} \nonumber \\ 
& f_{n} = \frac{1}{e^{\beta(\varepsilon_{n}-\mu)}+1},
\end{align}  
where $f_{n}$ is the Fermi distribution function. Unlike the zero temperature case, it is hard to make analytical progress in the finite-temperature case. So we numerically explore the behavior of persistent current at finite temperatures. In Fig.~\ref{fig4}, we show the behaviour of $I(\Phi)$ with $\gamma$ at finite temperatures for $\alpha=1.5$, $\Phi=0.25$. It is clear that at low temperature, the maximum persistent current occurs at small $\gamma$. But at higher temperatures, the maximum persistent current occurs at a finite filling fraction. To further investigate this effect, we calculate $\gamma|_{I_{max}}$ which is defined as the filling fraction at which the persistent current is maximum. The dependence of $\gamma|_{I_{max}}$ on various parameters is shown in the top panel of Fig.~\ref{fig5}.  The variation of $\gamma|_{I_{max}}$ with $\alpha$ is shown in Fig.~\ref{fig5}(a) for $N=100$, at various temperatures. At low temperature, ($T=0.05$), $\gamma|_{I_{max}}$ shows a jump with $\alpha$ from $\sim 1/N$ to a finite value. From the plot, it is seen that the transition point is shifted slightly from $\alpha=2$. This is due to finite-size effect. At higher temperature ($T=1.0$), $\gamma|_{I_{max}}$ smoothly increases with $\alpha$ and approaches a constant. So there is no phase transition, but instead a crossover. Fig.~\ref{fig5}(b), we show scaling of $\gamma|_{I_{max}}$ with $N$. It is clear from the figure that for $T=0.05$, $\gamma|_{I_{max}}\sim N^{-1}$ as long as Eq.~\ref{finite_temp_conditon} is satisfied. Thus, current is maximum for a sub-extensive number of particles. As system-size is increased further, $\gamma|_{I_{max}}$ tends to a constant, showing that current is maximum for an extensive number of particles. However, for a given $\alpha$ and given system size $N$, $\gamma|_{I_{max}}$ does not show a sharp change with increase in temperature, but rises smoothly  from being of $O(1/N)$ to a finite value. It can be seen from Fig.~\ref{fig5}(c).

In the bottom panel of Fig.~\ref{fig5} we show the behavior of the maximum persistent current $I_{max}$ with various parameters. Fig.~\ref{fig5}(d) shows plot of $I_{max}$ with $\alpha$ for various temperatures. As long as Eq.~\ref{finite_temp_conditon} is satisfied, $I_{max}$ decays exponentially with $\alpha$ with the same rate as at zero temperature. After that, $I_{max}$ decays exponentially, but at a faster rate, until it reaches close to the value corresponding to the nearest neighbour model. Then, it saturates to the value corresponding to the nearest neighbour model ($\alpha=\infty$). These effects are seen in Fig.~\ref{fig5}(d) for $T=0.05$. On the other hand, Eq.~\ref{finite_temp_conditon} is satisfied close to $\alpha\sim 1$ for $T=1.0$. Also, the maximum persistent current for the nearest neighbour model is exponentially small (see Appendix.~\ref{nnh_persistent}). As a result, for $T=1.0$, only the intermediate regime of exponential decay is seen for the choice of parameters. Fig.~\ref{fig5}(d) thus shows that the maximum persistent current for the power-law hopping case is exponentially larger than that of the nearest neighbour case ($\alpha=\infty$). Fig.~\ref{fig5}(e) shows plot of $I_{max}$ with $N$ for various values of temperature and $\alpha=1.7$. As long as Eq.~\ref{finite_temp_conditon} is satisfied, $I_{max}\sim N^{1-\alpha}$.  After that, it still decays as a power-law, but with a higher power-law exponent. This exponent depends on temperature. This is in contrast with the behavior for nearest neighbour case ($\alpha=\infty$), where it is known that the persistent current at a finite temperature decays exponentially with system size. Fig.~\ref{fig5}(f) shows plot of $I_{max}$ with $T$ for various values of $\alpha$ for $N=100$. Interestingly, it shows that $I_{max}$ decays as a power-law with power close to $\sim 2$. This is in stark contrast with that of the nearest neighbour model, where it is known that $I_{max}$ decays exponentially with temperature\cite{persistent2} (see Appendix.~\ref{nnh_persistent}). This is consistent with Fig.~\ref{fig5}(d). 

Thus, we have theoretically demonstrated that the effect of having maximum current at a sub-extensive number of particles due to power-law hopping can be experimentally observed by measuring persistent current in a mesoscopic ring. Further, the maximum persistent current at a finite temperature will be exponentially increased due to power-law hopping.

\subsection{Conclusion}

In this work, we have investigated the transport properties of 1D ordered number conserving fermionic systems with power-law hopping, the power-law exponent being given by $\alpha$.  Our most interesting result is that, in such systems, $1<\alpha<2$ describes a phase where the  zero temperature Drude weight diverges for a sub-extensive number of particles.  In other words, for $1<\alpha<2$, particle transport is maximum for a sub-extensive number of particles.   Since, for a sub-extensive number of particles there is negligible effect of interactions in the thermodynamic limit, the above statement is generally true in presence of arbitrary number conserving interaction terms.  We have also explicitly checked this for two  extremely different models of interaction. Thus, we have revealed very interesting universal transport behavior of 1D fermionic ordered systems with power-law hopping. This rather counter-intuitive effect  occurs because,  for $1<\alpha< 2$, the group velocity for the non-interacting system diverges at the bottom of the band. We have further shown that measurement of persistent current in a mesoscopic ring with power-law hopping will give experimental signature of this effect. For a mesoscopic system, this effect survives at a finite temperature. In fact, persistent current at a finite temperature is exponentially enhanced due to power-law hopping, as compared to that of nearest neighbour hopping. This observation can have interesting potential applications. 

Our results in this work opens a plethora of questions. This is more so because of an extreme lack of investigations related to transport properties of long-range systems. We have shown here that Drude weight diverges for a sub-extensive number of particles for generic ordered systems with power-law hopping. One interesting question is the effect of interactions on transport properties of systems with power-law hopping at a finite filling. This is extremely challenging to investigate because many numerical techniques fail for long-range systems. The combined effect of having finite-size effects up to large systems, the non-sparseness of the Hamiltonian matrix, the presence of long-range entanglement makes finite-size scaling from exact-diagonalization also difficult. We hope to rise to such challenges in future works.  

Another further interesting question is the effect of disorder or a quasi-periodic potential. In fact, it is known that there will be algebraic localization in such cases\cite{theory_power_law_loc5}. Understanding transport properties of such systems in the light of our results is of interest. Further, in this work, we have shown the consequence of non-analyticity of system's energy dispersion on the isolated system transport properties in thermodynamic limit.  Yet another extremely interesting question is the consequence of the non-analyticity in an open system transport set-up, i.e, where the central system is connected to two leads at different chemical potentials.  Detailed investigations in both these directions are currently on-going and will be published in subsequent works. 

\subsection{Acknowledgements}
M.S would like to acknowledge University Grants Commission (UGC) of India for her research fellowship and  SKM would like to acknowledge the financial support of DST-SERB, Government of India (Project File Number: EMR/2017/000504). A.P. acknowledges funding from the European Research Council (ERC) under the European Union’s Horizon 2020 research and innovation program (grant agreement No. 758403).

\appendix

\section{$\lim_{x\rightarrow 0^+} S_z(x)$}\label{Sz_limit}

Here we look at the behavior of $S_z(x)$ as small $x$, which is the most important mathematical result for governing our main result. The summation in the definition of $S_z(x)$ can be broken into two parts, one for $m\ll p/x$, and the other for $m\sim p/x$ and higher, where $p$ is an arbitrary number greater than $1$. Then, we have
\begin{align}
S_{z}(x)=\sum_{m=1}^{\infty} \frac{\sin(m x)}{m^z}=\sum_{m=1}^{p/x} \frac{\sin(m x)}{m^z}+x^z\sum_{y=p}^{\infty} \frac{\sin(y)}{y^z},
\end{align}
where, in the second term we have used $y=mx$. The summation in the second term is convergent, so in the limit $x\rightarrow 0^+$, this goes to zero for $z>0$. For small $x$, the first term can be evaluated as
\begin{align}
&S_z(x) \simeq x\sum_{m=1}^{p/x} \frac{1}{m^{z-1}}\simeq x\int_{m=1}^{p/x} dm \frac{1}{m^{z-1}} \nonumber \\
& =\frac{1}{2-z}\left(p^{2-z}x^{z-1}-1\right) ~~\forall z~\neq 1.
\end{align}
This shows that 
\begin{align}
\lim_{x\rightarrow 0^+} S_z(x) \sim \lim_{x\rightarrow 0^+} x^{z-1} ~~\forall z~\neq 1.
\end{align}
This is the result used in main text.

\section{$\lim_{\gamma\rightarrow 0^+} D(N)$}\label{DN_limit}
Here we check the $\lim_{\gamma\rightarrow 0^+} D(N)$ directly from Eq.~\ref{DN}. To check this, we rigorously convert the sum in Eq.~\ref{DN} into an integral by setting $x=m/N$, $dx=1/N$ and taking large $N$ limit. Then we have,
\begin{align}
D(N) &= 2N^{2-\alpha}\int\limits_{1/N}^{1/2} \frac{\sin[\pi N \gamma x]}{x^{\alpha -2}\sin[\pi x]} dx.             
\label{drudintegral}
\end{align}
For a single electron, $\gamma=1/N$, which tends to zero in the thermodynamic limit. For $\gamma=1/N$, from above equation,
\begin{align}
D(N) = \frac{2}{3-\alpha}\left(N^{2-\alpha}2^{\alpha-3}-\frac{1}{N}\right), ~~\forall~~\gamma=\frac{1}{N},~\alpha\neq 3
\end{align}
Thus, we have,
\begin{align}
\lim_{\gamma\rightarrow 0^+} D(N)  \sim
\begin{cases}
&  N^{2-\alpha},~\forall~1<\alpha\neq 2  \\
& 1,~\forall~\alpha=2 \\
\end{cases}.
\end{align}
Noting that for a sub-extensive number of particles, $\gamma\sim 1/N$, this is exactly consistent with the behavior from obtained from Eq.~\ref{D_analyt} and Eq.~\ref{Sz_result}.

\begin{figure}[h!]
\includegraphics[width=\columnwidth, height=5.5cm]{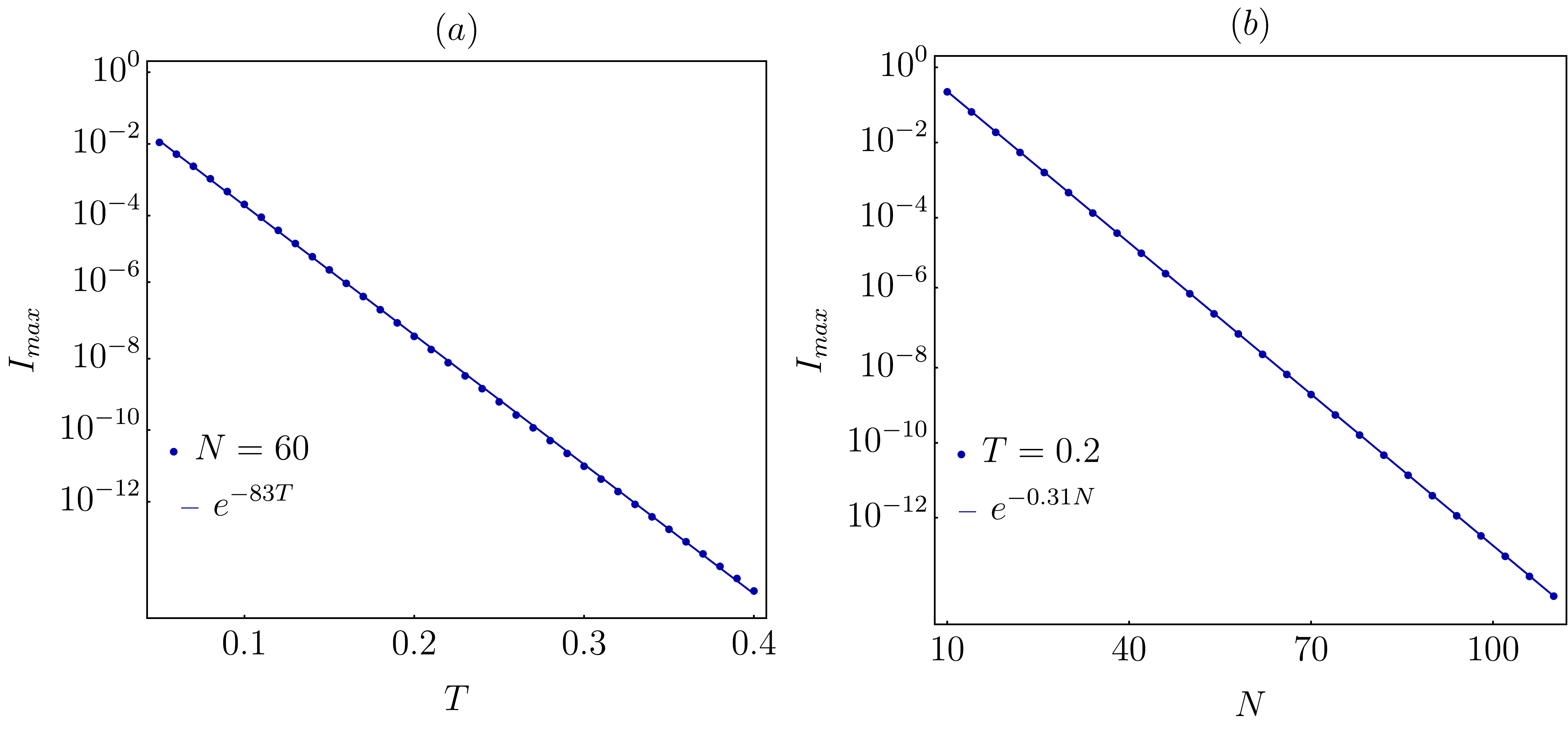}
\caption{(Color online) The figure shows variation the maximum persistent current for nearest neighbour non-interacting system with temperature and system-size. The maximum persistent current decays exponentially both with temperature and with system-size. } 
\label{fig:I_max_nn}
\end{figure}

\section{Nearest neighbour persistent current}\label{nnh_persistent}

Here we show that the persistent current for a non-interacting system with nearest neighbour hopping decays exponentially with temperature and with system size. This is shown by plotting the maximum persistent current with temperature and with system-size in a log-linear plot in Fig.~\ref{fig:I_max_nn}.

\bibliography{ref_long_range}

\begin{thebibliography}{85}%
\makeatletter
\providecommand \@ifxundefined [1]{%
 \@ifx{#1\undefined}
}%
\providecommand \@ifnum [1]{%
 \ifnum #1\expandafter \@firstoftwo
 \else \expandafter \@secondoftwo
 \fi
}%
\providecommand \@ifx [1]{%
 \ifx #1\expandafter \@firstoftwo
 \else \expandafter \@secondoftwo
 \fi
}%
\providecommand \natexlab [1]{#1}%
\providecommand \enquote  [1]{``#1''}%
\providecommand \bibnamefont  [1]{#1}%
\providecommand \bibfnamefont [1]{#1}%
\providecommand \citenamefont [1]{#1}%
\providecommand \href@noop [0]{\@secondoftwo}%
\providecommand \href [0]{\begingroup \@sanitize@url \@href}%
\providecommand \@href[1]{\@@startlink{#1}\@@href}%
\providecommand \@@href[1]{\endgroup#1\@@endlink}%
\providecommand \@sanitize@url [0]{\catcode `\\12\catcode `\$12\catcode
  `\&12\catcode `\#12\catcode `\^12\catcode `\_12\catcode `\%12\relax}%
\providecommand \@@startlink[1]{}%
\providecommand \@@endlink[0]{}%
\providecommand \url  [0]{\begingroup\@sanitize@url \@url }%
\providecommand \@url [1]{\endgroup\@href {#1}{\urlprefix }}%
\providecommand \urlprefix  [0]{URL }%
\providecommand \Eprint [0]{\href }%
\providecommand \doibase [0]{http://dx.doi.org/}%
\providecommand \selectlanguage [0]{\@gobble}%
\providecommand \bibinfo  [0]{\@secondoftwo}%
\providecommand \bibfield  [0]{\@secondoftwo}%
\providecommand \translation [1]{[#1]}%
\providecommand \BibitemOpen [0]{}%
\providecommand \bibitemStop [0]{}%
\providecommand \bibitemNoStop [0]{.\EOS\space}%
\providecommand \EOS [0]{\spacefactor3000\relax}%
\providecommand \BibitemShut  [1]{\csname bibitem#1\endcsname}%
\let\auto@bib@innerbib\@empty
\bibitem [{\citenamefont {Ryabtsev}\ \emph {et~al.}(2010)\citenamefont
  {Ryabtsev}, \citenamefont {Tretyakov}, \citenamefont {Beterov},\ and\
  \citenamefont {Entin}}]{expt_Rydberg_atoms1}%
  \BibitemOpen
  \bibfield  {author} {\bibinfo {author} {\bibfnamefont {I.~I.}\ \bibnamefont
  {Ryabtsev}}, \bibinfo {author} {\bibfnamefont {D.~B.}\ \bibnamefont
  {Tretyakov}}, \bibinfo {author} {\bibfnamefont {I.~I.}\ \bibnamefont
  {Beterov}}, \ and\ \bibinfo {author} {\bibfnamefont {V.~M.}\ \bibnamefont
  {Entin}},\ }\href {\doibase 10.1103/PhysRevLett.104.073003} {\bibfield
  {journal} {\bibinfo  {journal} {Phys. Rev. Lett.}\ }\textbf {\bibinfo
  {volume} {104}},\ \bibinfo {pages} {073003} (\bibinfo {year}
  {2010})}\BibitemShut {NoStop}%
\bibitem [{\citenamefont {B\'eguin}\ \emph {et~al.}(2013)\citenamefont
  {B\'eguin}, \citenamefont {Vernier}, \citenamefont {Chicireanu},
  \citenamefont {Lahaye},\ and\ \citenamefont
  {Browaeys}}]{expt_Rydberg_atoms2}%
  \BibitemOpen
  \bibfield  {author} {\bibinfo {author} {\bibfnamefont {L.}~\bibnamefont
  {B\'eguin}}, \bibinfo {author} {\bibfnamefont {A.}~\bibnamefont {Vernier}},
  \bibinfo {author} {\bibfnamefont {R.}~\bibnamefont {Chicireanu}}, \bibinfo
  {author} {\bibfnamefont {T.}~\bibnamefont {Lahaye}}, \ and\ \bibinfo {author}
  {\bibfnamefont {A.}~\bibnamefont {Browaeys}},\ }\href {\doibase
  10.1103/PhysRevLett.110.263201} {\bibfield  {journal} {\bibinfo  {journal}
  {Phys. Rev. Lett.}\ }\textbf {\bibinfo {volume} {110}},\ \bibinfo {pages}
  {263201} (\bibinfo {year} {2013})}\BibitemShut {NoStop}%
\bibitem [{\citenamefont {Browaeys}\ \emph {et~al.}(2016)\citenamefont
  {Browaeys}, \citenamefont {Barredo},\ and\ \citenamefont
  {Lahaye}}]{expt_Rydberg_atoms3}%
  \BibitemOpen
  \bibfield  {author} {\bibinfo {author} {\bibfnamefont {A.}~\bibnamefont
  {Browaeys}}, \bibinfo {author} {\bibfnamefont {D.}~\bibnamefont {Barredo}}, \
  and\ \bibinfo {author} {\bibfnamefont {T.}~\bibnamefont {Lahaye}},\ }\href
  {\doibase 10.1088/0953-4075/49/15/152001} {\bibfield  {journal} {\bibinfo
  {journal} {Journal of Physics B: Atomic, Molecular and Optical Physics}\
  }\textbf {\bibinfo {volume} {49}},\ \bibinfo {pages} {152001} (\bibinfo
  {year} {2016})}\BibitemShut {NoStop}%
\bibitem [{\citenamefont {Korenblit}\ \emph {et~al.}(2012)\citenamefont
  {Korenblit}, \citenamefont {Kafri}, \citenamefont {Campbell}, \citenamefont
  {Islam}, \citenamefont {Edwards}, \citenamefont {Gong}, \citenamefont {Lin},
  \citenamefont {Duan}, \citenamefont {Kim}, \citenamefont {Kim},\ and\
  \citenamefont {Monroe}}]{expt_trapped_ions1}%
  \BibitemOpen
  \bibfield  {author} {\bibinfo {author} {\bibfnamefont {S.}~\bibnamefont
  {Korenblit}}, \bibinfo {author} {\bibfnamefont {D.}~\bibnamefont {Kafri}},
  \bibinfo {author} {\bibfnamefont {W.~C.}\ \bibnamefont {Campbell}}, \bibinfo
  {author} {\bibfnamefont {R.}~\bibnamefont {Islam}}, \bibinfo {author}
  {\bibfnamefont {E.~E.}\ \bibnamefont {Edwards}}, \bibinfo {author}
  {\bibfnamefont {Z.-X.}\ \bibnamefont {Gong}}, \bibinfo {author}
  {\bibfnamefont {G.-D.}\ \bibnamefont {Lin}}, \bibinfo {author} {\bibfnamefont
  {L.-M.}\ \bibnamefont {Duan}}, \bibinfo {author} {\bibfnamefont
  {J.}~\bibnamefont {Kim}}, \bibinfo {author} {\bibfnamefont {K.}~\bibnamefont
  {Kim}}, \ and\ \bibinfo {author} {\bibfnamefont {C.}~\bibnamefont {Monroe}},\
  }\href {\doibase 10.1088/1367-2630/14/9/095024} {\bibfield  {journal}
  {\bibinfo  {journal} {New Journal of Physics}\ }\textbf {\bibinfo {volume}
  {14}},\ \bibinfo {pages} {095024} (\bibinfo {year} {2012})}\BibitemShut
  {NoStop}%
\bibitem [{\citenamefont {Jurcevic}\ \emph {et~al.}(2017)\citenamefont
  {Jurcevic}, \citenamefont {Shen}, \citenamefont {Hauke}, \citenamefont
  {Maier}, \citenamefont {Brydges}, \citenamefont {Hempel}, \citenamefont
  {Lanyon}, \citenamefont {Heyl}, \citenamefont {Blatt},\ and\ \citenamefont
  {Roos}}]{expt_trapped_ions2}%
  \BibitemOpen
  \bibfield  {author} {\bibinfo {author} {\bibfnamefont {P.}~\bibnamefont
  {Jurcevic}}, \bibinfo {author} {\bibfnamefont {H.}~\bibnamefont {Shen}},
  \bibinfo {author} {\bibfnamefont {P.}~\bibnamefont {Hauke}}, \bibinfo
  {author} {\bibfnamefont {C.}~\bibnamefont {Maier}}, \bibinfo {author}
  {\bibfnamefont {T.}~\bibnamefont {Brydges}}, \bibinfo {author} {\bibfnamefont
  {C.}~\bibnamefont {Hempel}}, \bibinfo {author} {\bibfnamefont {B.~P.}\
  \bibnamefont {Lanyon}}, \bibinfo {author} {\bibfnamefont {M.}~\bibnamefont
  {Heyl}}, \bibinfo {author} {\bibfnamefont {R.}~\bibnamefont {Blatt}}, \ and\
  \bibinfo {author} {\bibfnamefont {C.~F.}\ \bibnamefont {Roos}},\ }\href
  {\doibase 10.1103/PhysRevLett.119.080501} {\bibfield  {journal} {\bibinfo
  {journal} {Phys. Rev. Lett.}\ }\textbf {\bibinfo {volume} {119}},\ \bibinfo
  {pages} {080501} (\bibinfo {year} {2017})}\BibitemShut {NoStop}%
\bibitem [{\citenamefont {Neyenhuis}\ \emph {et~al.}(2017)\citenamefont
  {Neyenhuis}, \citenamefont {Zhang}, \citenamefont {Hess}, \citenamefont
  {Smith}, \citenamefont {Lee}, \citenamefont {Richerme}, \citenamefont {Gong},
  \citenamefont {Gorshkov},\ and\ \citenamefont {Monroe}}]{expt_trapped_ions3}%
  \BibitemOpen
  \bibfield  {author} {\bibinfo {author} {\bibfnamefont {B.}~\bibnamefont
  {Neyenhuis}}, \bibinfo {author} {\bibfnamefont {J.}~\bibnamefont {Zhang}},
  \bibinfo {author} {\bibfnamefont {P.~W.}\ \bibnamefont {Hess}}, \bibinfo
  {author} {\bibfnamefont {J.}~\bibnamefont {Smith}}, \bibinfo {author}
  {\bibfnamefont {A.~C.}\ \bibnamefont {Lee}}, \bibinfo {author} {\bibfnamefont
  {P.}~\bibnamefont {Richerme}}, \bibinfo {author} {\bibfnamefont {Z.-X.}\
  \bibnamefont {Gong}}, \bibinfo {author} {\bibfnamefont {A.~V.}\ \bibnamefont
  {Gorshkov}}, \ and\ \bibinfo {author} {\bibfnamefont {C.}~\bibnamefont
  {Monroe}},\ }\href {\doibase 10.1126/sciadv.1700672} {\bibfield  {journal}
  {\bibinfo  {journal} {Science Advances}\ }\textbf {\bibinfo {volume} {3}}
  (\bibinfo {year} {2017}),\ 10.1126/sciadv.1700672}\BibitemShut {NoStop}%
\bibitem [{\citenamefont {Britton}\ \emph {et~al.}(2012)\citenamefont
  {Britton}, \citenamefont {Sawyer}, \citenamefont {Keith}, \citenamefont
  {Wang}, \citenamefont {Freericks}, \citenamefont {Uys}, \citenamefont
  {Biercuk},\ and\ \citenamefont {Bollinger}}]{expt_trapped_ions4}%
  \BibitemOpen
  \bibfield  {author} {\bibinfo {author} {\bibfnamefont {J.~W.}\ \bibnamefont
  {Britton}}, \bibinfo {author} {\bibfnamefont {B.~C.}\ \bibnamefont {Sawyer}},
  \bibinfo {author} {\bibfnamefont {A.~C.}\ \bibnamefont {Keith}}, \bibinfo
  {author} {\bibfnamefont {C.~C.~J.}\ \bibnamefont {Wang}}, \bibinfo {author}
  {\bibfnamefont {J.~K.}\ \bibnamefont {Freericks}}, \bibinfo {author}
  {\bibfnamefont {H.}~\bibnamefont {Uys}}, \bibinfo {author} {\bibfnamefont
  {M.~J.}\ \bibnamefont {Biercuk}}, \ and\ \bibinfo {author} {\bibfnamefont
  {J.~J.}\ \bibnamefont {Bollinger}},\ }\href
  {https://doi.org/10.1038/nature10981} {\bibfield  {journal} {\bibinfo
  {journal} {Nature}\ }\textbf {\bibinfo {volume} {484}},\ \bibinfo {pages}
  {489 EP } (\bibinfo {year} {2012})}\BibitemShut {NoStop}%
\bibitem [{\citenamefont {Richerme}\ \emph {et~al.}(2014)\citenamefont
  {Richerme}, \citenamefont {Gong}, \citenamefont {Lee}, \citenamefont {Senko},
  \citenamefont {Smith}, \citenamefont {Foss-Feig}, \citenamefont {Michalakis},
  \citenamefont {Gorshkov},\ and\ \citenamefont {Monroe}}]{expt_trapped_ions5}%
  \BibitemOpen
  \bibfield  {author} {\bibinfo {author} {\bibfnamefont {P.}~\bibnamefont
  {Richerme}}, \bibinfo {author} {\bibfnamefont {Z.-X.}\ \bibnamefont {Gong}},
  \bibinfo {author} {\bibfnamefont {A.}~\bibnamefont {Lee}}, \bibinfo {author}
  {\bibfnamefont {C.}~\bibnamefont {Senko}}, \bibinfo {author} {\bibfnamefont
  {J.}~\bibnamefont {Smith}}, \bibinfo {author} {\bibfnamefont
  {M.}~\bibnamefont {Foss-Feig}}, \bibinfo {author} {\bibfnamefont
  {S.}~\bibnamefont {Michalakis}}, \bibinfo {author} {\bibfnamefont {A.~V.}\
  \bibnamefont {Gorshkov}}, \ and\ \bibinfo {author} {\bibfnamefont
  {C.}~\bibnamefont {Monroe}},\ }\href {https://doi.org/10.1038/nature13450}
  {\bibfield  {journal} {\bibinfo  {journal} {Nature}\ }\textbf {\bibinfo
  {volume} {511}},\ \bibinfo {pages} {198 EP } (\bibinfo {year}
  {2014})}\BibitemShut {NoStop}%
\bibitem [{\citenamefont {Jurcevic}\ \emph {et~al.}(2014)\citenamefont
  {Jurcevic}, \citenamefont {Lanyon}, \citenamefont {Hauke}, \citenamefont
  {Hempel}, \citenamefont {Zoller}, \citenamefont {Blatt},\ and\ \citenamefont
  {Roos}}]{expt_trapped_ions6}%
  \BibitemOpen
  \bibfield  {author} {\bibinfo {author} {\bibfnamefont {P.}~\bibnamefont
  {Jurcevic}}, \bibinfo {author} {\bibfnamefont {B.~P.}\ \bibnamefont
  {Lanyon}}, \bibinfo {author} {\bibfnamefont {P.}~\bibnamefont {Hauke}},
  \bibinfo {author} {\bibfnamefont {C.}~\bibnamefont {Hempel}}, \bibinfo
  {author} {\bibfnamefont {P.}~\bibnamefont {Zoller}}, \bibinfo {author}
  {\bibfnamefont {R.}~\bibnamefont {Blatt}}, \ and\ \bibinfo {author}
  {\bibfnamefont {C.~F.}\ \bibnamefont {Roos}},\ }\href
  {https://doi.org/10.1038/nature13461} {\bibfield  {journal} {\bibinfo
  {journal} {Nature}\ }\textbf {\bibinfo {volume} {511}},\ \bibinfo {pages}
  {202 EP } (\bibinfo {year} {2014})}\BibitemShut {NoStop}%
\bibitem [{\citenamefont {Zhang}\ \emph
  {et~al.}(2017{\natexlab{a}})\citenamefont {Zhang}, \citenamefont {Pagano},
  \citenamefont {Hess}, \citenamefont {Kyprianidis}, \citenamefont {Becker},
  \citenamefont {Kaplan}, \citenamefont {Gorshkov}, \citenamefont {Gong},\ and\
  \citenamefont {Monroe}}]{expt_trapped_ions7}%
  \BibitemOpen
  \bibfield  {author} {\bibinfo {author} {\bibfnamefont {J.}~\bibnamefont
  {Zhang}}, \bibinfo {author} {\bibfnamefont {G.}~\bibnamefont {Pagano}},
  \bibinfo {author} {\bibfnamefont {P.~W.}\ \bibnamefont {Hess}}, \bibinfo
  {author} {\bibfnamefont {A.}~\bibnamefont {Kyprianidis}}, \bibinfo {author}
  {\bibfnamefont {P.}~\bibnamefont {Becker}}, \bibinfo {author} {\bibfnamefont
  {H.}~\bibnamefont {Kaplan}}, \bibinfo {author} {\bibfnamefont {A.~V.}\
  \bibnamefont {Gorshkov}}, \bibinfo {author} {\bibfnamefont {Z.-X.}\
  \bibnamefont {Gong}}, \ and\ \bibinfo {author} {\bibfnamefont
  {C.}~\bibnamefont {Monroe}},\ }\href {https://doi.org/10.1038/nature24654}
  {\bibfield  {journal} {\bibinfo  {journal} {Nature}\ }\textbf {\bibinfo
  {volume} {551}},\ \bibinfo {pages} {601 EP } (\bibinfo {year}
  {2017}{\natexlab{a}})}\BibitemShut {NoStop}%
\bibitem [{\citenamefont {Zhang}\ \emph
  {et~al.}(2017{\natexlab{b}})\citenamefont {Zhang}, \citenamefont {Hess},
  \citenamefont {Kyprianidis}, \citenamefont {Becker}, \citenamefont {Lee},
  \citenamefont {Smith}, \citenamefont {Pagano}, \citenamefont {Potirniche},
  \citenamefont {Potter}, \citenamefont {Vishwanath}, \citenamefont {Yao},\
  and\ \citenamefont {Monroe}}]{expt_time_crystal2}%
  \BibitemOpen
  \bibfield  {author} {\bibinfo {author} {\bibfnamefont {J.}~\bibnamefont
  {Zhang}}, \bibinfo {author} {\bibfnamefont {P.~W.}\ \bibnamefont {Hess}},
  \bibinfo {author} {\bibfnamefont {A.}~\bibnamefont {Kyprianidis}}, \bibinfo
  {author} {\bibfnamefont {P.}~\bibnamefont {Becker}}, \bibinfo {author}
  {\bibfnamefont {A.}~\bibnamefont {Lee}}, \bibinfo {author} {\bibfnamefont
  {J.}~\bibnamefont {Smith}}, \bibinfo {author} {\bibfnamefont
  {G.}~\bibnamefont {Pagano}}, \bibinfo {author} {\bibfnamefont {I.-D.}\
  \bibnamefont {Potirniche}}, \bibinfo {author} {\bibfnamefont {A.~C.}\
  \bibnamefont {Potter}}, \bibinfo {author} {\bibfnamefont {A.}~\bibnamefont
  {Vishwanath}}, \bibinfo {author} {\bibfnamefont {N.~Y.}\ \bibnamefont {Yao}},
  \ and\ \bibinfo {author} {\bibfnamefont {C.}~\bibnamefont {Monroe}},\ }\href
  {https://doi.org/10.1038/nature21413} {\bibfield  {journal} {\bibinfo
  {journal} {Nature}\ }\textbf {\bibinfo {volume} {543}},\ \bibinfo {pages}
  {217 EP } (\bibinfo {year} {2017}{\natexlab{b}})}\BibitemShut {NoStop}%
\bibitem [{\citenamefont {Maier}\ \emph {et~al.}(2019)\citenamefont {Maier},
  \citenamefont {Brydges}, \citenamefont {Jurcevic}, \citenamefont {Trautmann},
  \citenamefont {Hempel}, \citenamefont {Lanyon}, \citenamefont {Hauke},
  \citenamefont {Blatt},\ and\ \citenamefont {Roos}}]{Experiment_transport}%
  \BibitemOpen
  \bibfield  {author} {\bibinfo {author} {\bibfnamefont {C.}~\bibnamefont
  {Maier}}, \bibinfo {author} {\bibfnamefont {T.}~\bibnamefont {Brydges}},
  \bibinfo {author} {\bibfnamefont {P.}~\bibnamefont {Jurcevic}}, \bibinfo
  {author} {\bibfnamefont {N.}~\bibnamefont {Trautmann}}, \bibinfo {author}
  {\bibfnamefont {C.}~\bibnamefont {Hempel}}, \bibinfo {author} {\bibfnamefont
  {B.~P.}\ \bibnamefont {Lanyon}}, \bibinfo {author} {\bibfnamefont
  {P.}~\bibnamefont {Hauke}}, \bibinfo {author} {\bibfnamefont
  {R.}~\bibnamefont {Blatt}}, \ and\ \bibinfo {author} {\bibfnamefont {C.~F.}\
  \bibnamefont {Roos}},\ }\href {\doibase 10.1103/PhysRevLett.122.050501}
  {\bibfield  {journal} {\bibinfo  {journal} {Phys. Rev. Lett.}\ }\textbf
  {\bibinfo {volume} {122}},\ \bibinfo {pages} {050501} (\bibinfo {year}
  {2019})}\BibitemShut {NoStop}%
\bibitem [{\citenamefont {Yan}\ \emph {et~al.}(2013)\citenamefont {Yan},
  \citenamefont {Moses}, \citenamefont {Gadway}, \citenamefont {Covey},
  \citenamefont {Hazzard}, \citenamefont {Rey}, \citenamefont {Jin},\ and\
  \citenamefont {Ye}}]{expt_polar_molecules1}%
  \BibitemOpen
  \bibfield  {author} {\bibinfo {author} {\bibfnamefont {B.}~\bibnamefont
  {Yan}}, \bibinfo {author} {\bibfnamefont {S.~A.}\ \bibnamefont {Moses}},
  \bibinfo {author} {\bibfnamefont {B.}~\bibnamefont {Gadway}}, \bibinfo
  {author} {\bibfnamefont {J.~P.}\ \bibnamefont {Covey}}, \bibinfo {author}
  {\bibfnamefont {K.~R.~A.}\ \bibnamefont {Hazzard}}, \bibinfo {author}
  {\bibfnamefont {A.~M.}\ \bibnamefont {Rey}}, \bibinfo {author} {\bibfnamefont
  {D.~S.}\ \bibnamefont {Jin}}, \ and\ \bibinfo {author} {\bibfnamefont
  {J.}~\bibnamefont {Ye}},\ }\href {https://doi.org/10.1038/nature12483}
  {\bibfield  {journal} {\bibinfo  {journal} {Nature}\ }\textbf {\bibinfo
  {volume} {501}},\ \bibinfo {pages} {521 EP } (\bibinfo {year}
  {2013})}\BibitemShut {NoStop}%
\bibitem [{\citenamefont {Ni}\ \emph {et~al.}(2008)\citenamefont {Ni},
  \citenamefont {Ospelkaus}, \citenamefont {de~Miranda}, \citenamefont
  {Pe{\textquoteright}er}, \citenamefont {Neyenhuis}, \citenamefont {Zirbel},
  \citenamefont {Kotochigova}, \citenamefont {Julienne}, \citenamefont {Jin},\
  and\ \citenamefont {Ye}}]{expt_polar_molecules2}%
  \BibitemOpen
  \bibfield  {author} {\bibinfo {author} {\bibfnamefont {K.-K.}\ \bibnamefont
  {Ni}}, \bibinfo {author} {\bibfnamefont {S.}~\bibnamefont {Ospelkaus}},
  \bibinfo {author} {\bibfnamefont {M.~H.~G.}\ \bibnamefont {de~Miranda}},
  \bibinfo {author} {\bibfnamefont {A.}~\bibnamefont {Pe{\textquoteright}er}},
  \bibinfo {author} {\bibfnamefont {B.}~\bibnamefont {Neyenhuis}}, \bibinfo
  {author} {\bibfnamefont {J.~J.}\ \bibnamefont {Zirbel}}, \bibinfo {author}
  {\bibfnamefont {S.}~\bibnamefont {Kotochigova}}, \bibinfo {author}
  {\bibfnamefont {P.~S.}\ \bibnamefont {Julienne}}, \bibinfo {author}
  {\bibfnamefont {D.~S.}\ \bibnamefont {Jin}}, \ and\ \bibinfo {author}
  {\bibfnamefont {J.}~\bibnamefont {Ye}},\ }\href {\doibase
  10.1126/science.1163861} {\bibfield  {journal} {\bibinfo  {journal}
  {Science}\ }\textbf {\bibinfo {volume} {322}},\ \bibinfo {pages} {231}
  (\bibinfo {year} {2008})}\BibitemShut {NoStop}%
\bibitem [{\citenamefont {Moses}\ \emph {et~al.}(2016)\citenamefont {Moses},
  \citenamefont {Covey}, \citenamefont {Miecnikowski}, \citenamefont {Jin},\
  and\ \citenamefont {Ye}}]{expt_polar_molecules3}%
  \BibitemOpen
  \bibfield  {author} {\bibinfo {author} {\bibfnamefont {S.~A.}\ \bibnamefont
  {Moses}}, \bibinfo {author} {\bibfnamefont {J.~P.}\ \bibnamefont {Covey}},
  \bibinfo {author} {\bibfnamefont {M.~T.}\ \bibnamefont {Miecnikowski}},
  \bibinfo {author} {\bibfnamefont {D.~S.}\ \bibnamefont {Jin}}, \ and\
  \bibinfo {author} {\bibfnamefont {J.}~\bibnamefont {Ye}},\ }\href
  {https://doi.org/10.1038/nphys3985} {\bibfield  {journal} {\bibinfo
  {journal} {Nature Physics}\ }\textbf {\bibinfo {volume} {13}},\ \bibinfo
  {pages} {13 EP } (\bibinfo {year} {2016})}\BibitemShut {NoStop}%
\bibitem [{\citenamefont {de~Paz}\ \emph {et~al.}(2013)\citenamefont {de~Paz},
  \citenamefont {Sharma}, \citenamefont {Chotia}, \citenamefont {Mar\'echal},
  \citenamefont {Huckans}, \citenamefont {Pedri}, \citenamefont {Santos},
  \citenamefont {Gorceix}, \citenamefont {Vernac},\ and\ \citenamefont
  {Laburthe-Tolra}}]{expt_dipolar_gas1}%
  \BibitemOpen
  \bibfield  {author} {\bibinfo {author} {\bibfnamefont {A.}~\bibnamefont
  {de~Paz}}, \bibinfo {author} {\bibfnamefont {A.}~\bibnamefont {Sharma}},
  \bibinfo {author} {\bibfnamefont {A.}~\bibnamefont {Chotia}}, \bibinfo
  {author} {\bibfnamefont {E.}~\bibnamefont {Mar\'echal}}, \bibinfo {author}
  {\bibfnamefont {J.~H.}\ \bibnamefont {Huckans}}, \bibinfo {author}
  {\bibfnamefont {P.}~\bibnamefont {Pedri}}, \bibinfo {author} {\bibfnamefont
  {L.}~\bibnamefont {Santos}}, \bibinfo {author} {\bibfnamefont
  {O.}~\bibnamefont {Gorceix}}, \bibinfo {author} {\bibfnamefont
  {L.}~\bibnamefont {Vernac}}, \ and\ \bibinfo {author} {\bibfnamefont
  {B.}~\bibnamefont {Laburthe-Tolra}},\ }\href {\doibase
  10.1103/PhysRevLett.111.185305} {\bibfield  {journal} {\bibinfo  {journal}
  {Phys. Rev. Lett.}\ }\textbf {\bibinfo {volume} {111}},\ \bibinfo {pages}
  {185305} (\bibinfo {year} {2013})}\BibitemShut {NoStop}%
\bibitem [{\citenamefont {{\'A}lvarez}\ \emph {et~al.}(2015)\citenamefont
  {{\'A}lvarez}, \citenamefont {Suter},\ and\ \citenamefont
  {Kaiser}}]{expt_nuclear_spins}%
  \BibitemOpen
  \bibfield  {author} {\bibinfo {author} {\bibfnamefont {G.~A.}\ \bibnamefont
  {{\'A}lvarez}}, \bibinfo {author} {\bibfnamefont {D.}~\bibnamefont {Suter}},
  \ and\ \bibinfo {author} {\bibfnamefont {R.}~\bibnamefont {Kaiser}},\ }\href
  {\doibase 10.1126/science.1261160} {\bibfield  {journal} {\bibinfo  {journal}
  {Science}\ }\textbf {\bibinfo {volume} {349}},\ \bibinfo {pages} {846}
  (\bibinfo {year} {2015})}\BibitemShut {NoStop}%
\bibitem [{\citenamefont {Choi}\ \emph {et~al.}(2017)\citenamefont {Choi},
  \citenamefont {Choi}, \citenamefont {Landig}, \citenamefont {Kucsko},
  \citenamefont {Zhou}, \citenamefont {Isoya}, \citenamefont {Jelezko},
  \citenamefont {Onoda}, \citenamefont {Sumiya}, \citenamefont {Khemani},
  \citenamefont {von Keyserlingk}, \citenamefont {Yao}, \citenamefont
  {Demler},\ and\ \citenamefont {Lukin}}]{expt_time_crystal1}%
  \BibitemOpen
  \bibfield  {author} {\bibinfo {author} {\bibfnamefont {S.}~\bibnamefont
  {Choi}}, \bibinfo {author} {\bibfnamefont {J.}~\bibnamefont {Choi}}, \bibinfo
  {author} {\bibfnamefont {R.}~\bibnamefont {Landig}}, \bibinfo {author}
  {\bibfnamefont {G.}~\bibnamefont {Kucsko}}, \bibinfo {author} {\bibfnamefont
  {H.}~\bibnamefont {Zhou}}, \bibinfo {author} {\bibfnamefont {J.}~\bibnamefont
  {Isoya}}, \bibinfo {author} {\bibfnamefont {F.}~\bibnamefont {Jelezko}},
  \bibinfo {author} {\bibfnamefont {S.}~\bibnamefont {Onoda}}, \bibinfo
  {author} {\bibfnamefont {H.}~\bibnamefont {Sumiya}}, \bibinfo {author}
  {\bibfnamefont {V.}~\bibnamefont {Khemani}}, \bibinfo {author} {\bibfnamefont
  {C.}~\bibnamefont {von Keyserlingk}}, \bibinfo {author} {\bibfnamefont
  {N.~Y.}\ \bibnamefont {Yao}}, \bibinfo {author} {\bibfnamefont
  {E.}~\bibnamefont {Demler}}, \ and\ \bibinfo {author} {\bibfnamefont {M.~D.}\
  \bibnamefont {Lukin}},\ }\href {https://doi.org/10.1038/nature21426}
  {\bibfield  {journal} {\bibinfo  {journal} {Nature}\ }\textbf {\bibinfo
  {volume} {543}},\ \bibinfo {pages} {221 EP } (\bibinfo {year}
  {2017})}\BibitemShut {NoStop}%
\bibitem [{\citenamefont {Hung}\ \emph {et~al.}(2016)\citenamefont {Hung},
  \citenamefont {Gonz{\'a}lez-Tudela}, \citenamefont {Cirac},\ and\
  \citenamefont {Kimble}}]{expt_cold_atoms_in_waveguides1}%
  \BibitemOpen
  \bibfield  {author} {\bibinfo {author} {\bibfnamefont {C.-L.}\ \bibnamefont
  {Hung}}, \bibinfo {author} {\bibfnamefont {A.}~\bibnamefont
  {Gonz{\'a}lez-Tudela}}, \bibinfo {author} {\bibfnamefont {J.~I.}\
  \bibnamefont {Cirac}}, \ and\ \bibinfo {author} {\bibfnamefont {H.~J.}\
  \bibnamefont {Kimble}},\ }\href {\doibase 10.1073/pnas.1603777113} {\bibfield
   {journal} {\bibinfo  {journal} {Proceedings of the National Academy of
  Sciences}\ }\textbf {\bibinfo {volume} {113}},\ \bibinfo {pages} {E4946}
  (\bibinfo {year} {2016})}\BibitemShut {NoStop}%
\bibitem [{\citenamefont {Smale}\ \emph {et~al.}(2019)\citenamefont {Smale},
  \citenamefont {He}, \citenamefont {Olsen}, \citenamefont {Jackson},
  \citenamefont {Sharum}, \citenamefont {Trotzky}, \citenamefont {Marino},
  \citenamefont {Rey},\ and\ \citenamefont {Thywissen}}]{expt_atoms_in_trap}%
  \BibitemOpen
  \bibfield  {author} {\bibinfo {author} {\bibfnamefont {S.}~\bibnamefont
  {Smale}}, \bibinfo {author} {\bibfnamefont {P.}~\bibnamefont {He}}, \bibinfo
  {author} {\bibfnamefont {B.~A.}\ \bibnamefont {Olsen}}, \bibinfo {author}
  {\bibfnamefont {K.~G.}\ \bibnamefont {Jackson}}, \bibinfo {author}
  {\bibfnamefont {H.}~\bibnamefont {Sharum}}, \bibinfo {author} {\bibfnamefont
  {S.}~\bibnamefont {Trotzky}}, \bibinfo {author} {\bibfnamefont
  {J.}~\bibnamefont {Marino}}, \bibinfo {author} {\bibfnamefont {A.~M.}\
  \bibnamefont {Rey}}, \ and\ \bibinfo {author} {\bibfnamefont {J.~H.}\
  \bibnamefont {Thywissen}},\ }\href {\doibase 10.1126/sciadv.aax1568}
  {\bibfield  {journal} {\bibinfo  {journal} {Science Advances}\ }\textbf
  {\bibinfo {volume} {5}} (\bibinfo {year} {2019}),\ 10.1126/sciadv.aax1568},\
  \Eprint
  {http://arxiv.org/abs/https://advances.sciencemag.org/content/5/8/eaax1568.full.pdf}
  {https://advances.sciencemag.org/content/5/8/eaax1568.full.pdf} \BibitemShut
  {NoStop}%
\bibitem [{\citenamefont {\ifmmode \check{Z}\else
  \v{Z}\fi{}unkovi\ifmmode~\check{c}\else \v{c}\fi{}}\ \emph
  {et~al.}(2018)\citenamefont {\ifmmode \check{Z}\else
  \v{Z}\fi{}unkovi\ifmmode~\check{c}\else \v{c}\fi{}}, \citenamefont {Heyl},
  \citenamefont {Knap},\ and\ \citenamefont {Silva}}]{theory_DQPT1}%
  \BibitemOpen
  \bibfield  {author} {\bibinfo {author} {\bibfnamefont {B.}~\bibnamefont
  {\ifmmode \check{Z}\else \v{Z}\fi{}unkovi\ifmmode~\check{c}\else
  \v{c}\fi{}}}, \bibinfo {author} {\bibfnamefont {M.}~\bibnamefont {Heyl}},
  \bibinfo {author} {\bibfnamefont {M.}~\bibnamefont {Knap}}, \ and\ \bibinfo
  {author} {\bibfnamefont {A.}~\bibnamefont {Silva}},\ }\href {\doibase
  10.1103/PhysRevLett.120.130601} {\bibfield  {journal} {\bibinfo  {journal}
  {Phys. Rev. Lett.}\ }\textbf {\bibinfo {volume} {120}},\ \bibinfo {pages}
  {130601} (\bibinfo {year} {2018})}\BibitemShut {NoStop}%
\bibitem [{\citenamefont {Halimeh}\ and\ \citenamefont
  {Zauner-Stauber}(2017)}]{theory_DQPT2}%
  \BibitemOpen
  \bibfield  {author} {\bibinfo {author} {\bibfnamefont {J.~C.}\ \bibnamefont
  {Halimeh}}\ and\ \bibinfo {author} {\bibfnamefont {V.}~\bibnamefont
  {Zauner-Stauber}},\ }\href {\doibase 10.1103/PhysRevB.96.134427} {\bibfield
  {journal} {\bibinfo  {journal} {Phys. Rev. B}\ }\textbf {\bibinfo {volume}
  {96}},\ \bibinfo {pages} {134427} (\bibinfo {year} {2017})}\BibitemShut
  {NoStop}%
\bibitem [{\citenamefont {Gong}\ and\ \citenamefont
  {Duan}(2013)}]{theory_pretherm1}%
  \BibitemOpen
  \bibfield  {author} {\bibinfo {author} {\bibfnamefont {Z.-X.}\ \bibnamefont
  {Gong}}\ and\ \bibinfo {author} {\bibfnamefont {L.-M.}\ \bibnamefont
  {Duan}},\ }\href {\doibase 10.1088/1367-2630/15/11/113051} {\bibfield
  {journal} {\bibinfo  {journal} {New Journal of Physics}\ }\textbf {\bibinfo
  {volume} {15}},\ \bibinfo {pages} {113051} (\bibinfo {year}
  {2013})}\BibitemShut {NoStop}%
\bibitem [{\citenamefont {Mori}(2019)}]{theory_pretherm2}%
  \BibitemOpen
  \bibfield  {author} {\bibinfo {author} {\bibfnamefont {T.}~\bibnamefont
  {Mori}},\ }\href {\doibase 10.1088/1751-8121/aaf9db} {\bibfield  {journal}
  {\bibinfo  {journal} {Journal of Physics A: Mathematical and Theoretical}\
  }\textbf {\bibinfo {volume} {52}},\ \bibinfo {pages} {054001} (\bibinfo
  {year} {2019})}\BibitemShut {NoStop}%
\bibitem [{\citenamefont {Lerose}\ \emph {et~al.}(2019)\citenamefont {Lerose},
  \citenamefont {\ifmmode \check{Z}\else
  \v{Z}\fi{}unkovi\ifmmode~\check{c}\else \v{c}\fi{}}, \citenamefont {Silva},\
  and\ \citenamefont {Gambassi}}]{theory_pretherm3}%
  \BibitemOpen
  \bibfield  {author} {\bibinfo {author} {\bibfnamefont {A.}~\bibnamefont
  {Lerose}}, \bibinfo {author} {\bibfnamefont {B.}~\bibnamefont {\ifmmode
  \check{Z}\else \v{Z}\fi{}unkovi\ifmmode~\check{c}\else \v{c}\fi{}}}, \bibinfo
  {author} {\bibfnamefont {A.}~\bibnamefont {Silva}}, \ and\ \bibinfo {author}
  {\bibfnamefont {A.}~\bibnamefont {Gambassi}},\ }\href {\doibase
  10.1103/PhysRevB.99.121112} {\bibfield  {journal} {\bibinfo  {journal} {Phys.
  Rev. B}\ }\textbf {\bibinfo {volume} {99}},\ \bibinfo {pages} {121112}
  (\bibinfo {year} {2019})}\BibitemShut {NoStop}%
\bibitem [{\citenamefont {Dutta}\ and\ \citenamefont
  {Bhattacharjee}(2001)}]{theory_power_law14}%
  \BibitemOpen
  \bibfield  {author} {\bibinfo {author} {\bibfnamefont {A.}~\bibnamefont
  {Dutta}}\ and\ \bibinfo {author} {\bibfnamefont {J.~K.}\ \bibnamefont
  {Bhattacharjee}},\ }\href {\doibase 10.1103/PhysRevB.64.184106} {\bibfield
  {journal} {\bibinfo  {journal} {Phys. Rev. B}\ }\textbf {\bibinfo {volume}
  {64}},\ \bibinfo {pages} {184106} (\bibinfo {year} {2001})}\BibitemShut
  {NoStop}%
\bibitem [{\citenamefont {Dutta}\ and\ \citenamefont
  {Dutta}(2017)}]{theory_power_law15}%
  \BibitemOpen
  \bibfield  {author} {\bibinfo {author} {\bibfnamefont {A.}~\bibnamefont
  {Dutta}}\ and\ \bibinfo {author} {\bibfnamefont {A.}~\bibnamefont {Dutta}},\
  }\href {\doibase 10.1103/PhysRevB.96.125113} {\bibfield  {journal} {\bibinfo
  {journal} {Phys. Rev. B}\ }\textbf {\bibinfo {volume} {96}},\ \bibinfo
  {pages} {125113} (\bibinfo {year} {2017})}\BibitemShut {NoStop}%
\bibitem [{\citenamefont {Lerose}\ \emph
  {et~al.}(2018{\natexlab{a}})\citenamefont {Lerose}, \citenamefont {Marino},
  \citenamefont {Gambassi},\ and\ \citenamefont
  {Silva}}]{alessio_kaptiza_2018}%
  \BibitemOpen
  \bibfield  {author} {\bibinfo {author} {\bibfnamefont {A.}~\bibnamefont
  {Lerose}}, \bibinfo {author} {\bibfnamefont {J.}~\bibnamefont {Marino}},
  \bibinfo {author} {\bibfnamefont {A.}~\bibnamefont {Gambassi}}, \ and\
  \bibinfo {author} {\bibfnamefont {A.}~\bibnamefont {Silva}},\ }\href
  {https://arxiv.org/abs/1803.04490} {\bibfield  {journal} {\bibinfo  {journal}
  {arXiv:1803.04490}\ } (\bibinfo {year} {2018}{\natexlab{a}})}\BibitemShut
  {NoStop}%
\bibitem [{\citenamefont {Hauke}\ and\ \citenamefont
  {Tagliacozzo}(2013)}]{theory_LRB1}%
  \BibitemOpen
  \bibfield  {author} {\bibinfo {author} {\bibfnamefont {P.}~\bibnamefont
  {Hauke}}\ and\ \bibinfo {author} {\bibfnamefont {L.}~\bibnamefont
  {Tagliacozzo}},\ }\href {\doibase 10.1103/PhysRevLett.111.207202} {\bibfield
  {journal} {\bibinfo  {journal} {Phys. Rev. Lett.}\ }\textbf {\bibinfo
  {volume} {111}},\ \bibinfo {pages} {207202} (\bibinfo {year}
  {2013})}\BibitemShut {NoStop}%
\bibitem [{\citenamefont {Eisert}\ \emph {et~al.}(2013)\citenamefont {Eisert},
  \citenamefont {van~den Worm}, \citenamefont {Manmana},\ and\ \citenamefont
  {Kastner}}]{theory_LRB2}%
  \BibitemOpen
  \bibfield  {author} {\bibinfo {author} {\bibfnamefont {J.}~\bibnamefont
  {Eisert}}, \bibinfo {author} {\bibfnamefont {M.}~\bibnamefont {van~den
  Worm}}, \bibinfo {author} {\bibfnamefont {S.~R.}\ \bibnamefont {Manmana}}, \
  and\ \bibinfo {author} {\bibfnamefont {M.}~\bibnamefont {Kastner}},\ }\href
  {\doibase 10.1103/PhysRevLett.111.260401} {\bibfield  {journal} {\bibinfo
  {journal} {Phys. Rev. Lett.}\ }\textbf {\bibinfo {volume} {111}},\ \bibinfo
  {pages} {260401} (\bibinfo {year} {2013})}\BibitemShut {NoStop}%
\bibitem [{\citenamefont {Gong}\ \emph {et~al.}(2014)\citenamefont {Gong},
  \citenamefont {Foss-Feig}, \citenamefont {Michalakis},\ and\ \citenamefont
  {Gorshkov}}]{theory_LRB3}%
  \BibitemOpen
  \bibfield  {author} {\bibinfo {author} {\bibfnamefont {Z.-X.}\ \bibnamefont
  {Gong}}, \bibinfo {author} {\bibfnamefont {M.}~\bibnamefont {Foss-Feig}},
  \bibinfo {author} {\bibfnamefont {S.}~\bibnamefont {Michalakis}}, \ and\
  \bibinfo {author} {\bibfnamefont {A.~V.}\ \bibnamefont {Gorshkov}},\ }\href
  {\doibase 10.1103/PhysRevLett.113.030602} {\bibfield  {journal} {\bibinfo
  {journal} {Phys. Rev. Lett.}\ }\textbf {\bibinfo {volume} {113}},\ \bibinfo
  {pages} {030602} (\bibinfo {year} {2014})}\BibitemShut {NoStop}%
\bibitem [{\citenamefont {Foss-Feig}\ \emph {et~al.}(2015)\citenamefont
  {Foss-Feig}, \citenamefont {Gong}, \citenamefont {Clark},\ and\ \citenamefont
  {Gorshkov}}]{theory_LRB4}%
  \BibitemOpen
  \bibfield  {author} {\bibinfo {author} {\bibfnamefont {M.}~\bibnamefont
  {Foss-Feig}}, \bibinfo {author} {\bibfnamefont {Z.-X.}\ \bibnamefont {Gong}},
  \bibinfo {author} {\bibfnamefont {C.~W.}\ \bibnamefont {Clark}}, \ and\
  \bibinfo {author} {\bibfnamefont {A.~V.}\ \bibnamefont {Gorshkov}},\ }\href
  {\doibase 10.1103/PhysRevLett.114.157201} {\bibfield  {journal} {\bibinfo
  {journal} {Phys. Rev. Lett.}\ }\textbf {\bibinfo {volume} {114}},\ \bibinfo
  {pages} {157201} (\bibinfo {year} {2015})}\BibitemShut {NoStop}%
\bibitem [{\citenamefont {Cevolani}\ \emph {et~al.}(2015)\citenamefont
  {Cevolani}, \citenamefont {Carleo},\ and\ \citenamefont
  {Sanchez-Palencia}}]{theory_LRB5}%
  \BibitemOpen
  \bibfield  {author} {\bibinfo {author} {\bibfnamefont {L.}~\bibnamefont
  {Cevolani}}, \bibinfo {author} {\bibfnamefont {G.}~\bibnamefont {Carleo}}, \
  and\ \bibinfo {author} {\bibfnamefont {L.}~\bibnamefont {Sanchez-Palencia}},\
  }\href {\doibase 10.1103/PhysRevA.92.041603} {\bibfield  {journal} {\bibinfo
  {journal} {Phys. Rev. A}\ }\textbf {\bibinfo {volume} {92}},\ \bibinfo
  {pages} {041603} (\bibinfo {year} {2015})}\BibitemShut {NoStop}%
\bibitem [{\citenamefont {Cevolani}\ \emph {et~al.}(2016)\citenamefont
  {Cevolani}, \citenamefont {Carleo},\ and\ \citenamefont
  {Sanchez-Palencia}}]{theory_LRB6}%
  \BibitemOpen
  \bibfield  {author} {\bibinfo {author} {\bibfnamefont {L.}~\bibnamefont
  {Cevolani}}, \bibinfo {author} {\bibfnamefont {G.}~\bibnamefont {Carleo}}, \
  and\ \bibinfo {author} {\bibfnamefont {L.}~\bibnamefont {Sanchez-Palencia}},\
  }\href {\doibase 10.1088/1367-2630/18/9/093002} {\bibfield  {journal}
  {\bibinfo  {journal} {New Journal of Physics}\ }\textbf {\bibinfo {volume}
  {18}},\ \bibinfo {pages} {093002} (\bibinfo {year} {2016})}\BibitemShut
  {NoStop}%
\bibitem [{\citenamefont {Fr\'erot}\ \emph {et~al.}(2018)\citenamefont
  {Fr\'erot}, \citenamefont {Naldesi},\ and\ \citenamefont
  {Roscilde}}]{theory_LRB7}%
  \BibitemOpen
  \bibfield  {author} {\bibinfo {author} {\bibfnamefont {I.}~\bibnamefont
  {Fr\'erot}}, \bibinfo {author} {\bibfnamefont {P.}~\bibnamefont {Naldesi}}, \
  and\ \bibinfo {author} {\bibfnamefont {T.}~\bibnamefont {Roscilde}},\ }\href
  {\doibase 10.1103/PhysRevLett.120.050401} {\bibfield  {journal} {\bibinfo
  {journal} {Phys. Rev. Lett.}\ }\textbf {\bibinfo {volume} {120}},\ \bibinfo
  {pages} {050401} (\bibinfo {year} {2018})}\BibitemShut {NoStop}%
\bibitem [{\citenamefont {Tran}\ \emph {et~al.}(2019)\citenamefont {Tran},
  \citenamefont {Guo}, \citenamefont {Su}, \citenamefont {Garrison},
  \citenamefont {Eldredge}, \citenamefont {Foss-Feig}, \citenamefont {Childs},\
  and\ \citenamefont {Gorshkov}}]{theory_LRB8}%
  \BibitemOpen
  \bibfield  {author} {\bibinfo {author} {\bibfnamefont {M.~C.}\ \bibnamefont
  {Tran}}, \bibinfo {author} {\bibfnamefont {A.~Y.}\ \bibnamefont {Guo}},
  \bibinfo {author} {\bibfnamefont {Y.}~\bibnamefont {Su}}, \bibinfo {author}
  {\bibfnamefont {J.~R.}\ \bibnamefont {Garrison}}, \bibinfo {author}
  {\bibfnamefont {Z.}~\bibnamefont {Eldredge}}, \bibinfo {author}
  {\bibfnamefont {M.}~\bibnamefont {Foss-Feig}}, \bibinfo {author}
  {\bibfnamefont {A.~M.}\ \bibnamefont {Childs}}, \ and\ \bibinfo {author}
  {\bibfnamefont {A.~V.}\ \bibnamefont {Gorshkov}},\ }\href {\doibase
  10.1103/PhysRevX.9.031006} {\bibfield  {journal} {\bibinfo  {journal} {Phys.
  Rev. X}\ }\textbf {\bibinfo {volume} {9}},\ \bibinfo {pages} {031006}
  (\bibinfo {year} {2019})}\BibitemShut {NoStop}%
\bibitem [{\citenamefont {Luitz}\ and\ \citenamefont
  {Bar~Lev}(2019)}]{theory_LRB9}%
  \BibitemOpen
  \bibfield  {author} {\bibinfo {author} {\bibfnamefont {D.~J.}\ \bibnamefont
  {Luitz}}\ and\ \bibinfo {author} {\bibfnamefont {Y.}~\bibnamefont
  {Bar~Lev}},\ }\href {\doibase 10.1103/PhysRevA.99.010105} {\bibfield
  {journal} {\bibinfo  {journal} {Phys. Rev. A}\ }\textbf {\bibinfo {volume}
  {99}},\ \bibinfo {pages} {010105} (\bibinfo {year} {2019})}\BibitemShut
  {NoStop}%
\bibitem [{\citenamefont {Cevolani}\ \emph {et~al.}(2018)\citenamefont
  {Cevolani}, \citenamefont {Despres}, \citenamefont {Carleo}, \citenamefont
  {Tagliacozzo},\ and\ \citenamefont {Sanchez-Palencia}}]{Laurent}%
  \BibitemOpen
  \bibfield  {author} {\bibinfo {author} {\bibfnamefont {L.}~\bibnamefont
  {Cevolani}}, \bibinfo {author} {\bibfnamefont {J.}~\bibnamefont {Despres}},
  \bibinfo {author} {\bibfnamefont {G.}~\bibnamefont {Carleo}}, \bibinfo
  {author} {\bibfnamefont {L.}~\bibnamefont {Tagliacozzo}}, \ and\ \bibinfo
  {author} {\bibfnamefont {L.}~\bibnamefont {Sanchez-Palencia}},\ }\href
  {\doibase 10.1103/PhysRevB.98.024302} {\bibfield  {journal} {\bibinfo
  {journal} {Phys. Rev. B}\ }\textbf {\bibinfo {volume} {98}},\ \bibinfo
  {pages} {024302} (\bibinfo {year} {2018})}\BibitemShut {NoStop}%
\bibitem [{\citenamefont {{Modak}}\ and\ \citenamefont
  {{Nag}}(2019)}]{Ranjan_Tanay}%
  \BibitemOpen
  \bibfield  {author} {\bibinfo {author} {\bibfnamefont {R.}~\bibnamefont
  {{Modak}}}\ and\ \bibinfo {author} {\bibfnamefont {T.}~\bibnamefont
  {{Nag}}},\ }\href@noop {} {\bibfield  {journal} {\bibinfo  {journal} {arXiv
  e-prints}\ ,\ \bibinfo {eid} {arXiv:1903.05099}} (\bibinfo {year} {2019})},\
  \Eprint {http://arxiv.org/abs/1903.05099} {arXiv:1903.05099
  [cond-mat.dis-nn]} \BibitemShut {NoStop}%
\bibitem [{\citenamefont {Koffel}\ \emph {et~al.}(2012)\citenamefont {Koffel},
  \citenamefont {Lewenstein},\ and\ \citenamefont
  {Tagliacozzo}}]{theory_entanglement1}%
  \BibitemOpen
  \bibfield  {author} {\bibinfo {author} {\bibfnamefont {T.}~\bibnamefont
  {Koffel}}, \bibinfo {author} {\bibfnamefont {M.}~\bibnamefont {Lewenstein}},
  \ and\ \bibinfo {author} {\bibfnamefont {L.}~\bibnamefont {Tagliacozzo}},\
  }\href {\doibase 10.1103/PhysRevLett.109.267203} {\bibfield  {journal}
  {\bibinfo  {journal} {Phys. Rev. Lett.}\ }\textbf {\bibinfo {volume} {109}},\
  \bibinfo {pages} {267203} (\bibinfo {year} {2012})}\BibitemShut {NoStop}%
\bibitem [{\citenamefont {Vodola}\ \emph {et~al.}(2014)\citenamefont {Vodola},
  \citenamefont {Lepori}, \citenamefont {Ercolessi}, \citenamefont {Gorshkov},\
  and\ \citenamefont {Pupillo}}]{theory_entanglement2}%
  \BibitemOpen
  \bibfield  {author} {\bibinfo {author} {\bibfnamefont {D.}~\bibnamefont
  {Vodola}}, \bibinfo {author} {\bibfnamefont {L.}~\bibnamefont {Lepori}},
  \bibinfo {author} {\bibfnamefont {E.}~\bibnamefont {Ercolessi}}, \bibinfo
  {author} {\bibfnamefont {A.~V.}\ \bibnamefont {Gorshkov}}, \ and\ \bibinfo
  {author} {\bibfnamefont {G.}~\bibnamefont {Pupillo}},\ }\href {\doibase
  10.1103/PhysRevLett.113.156402} {\bibfield  {journal} {\bibinfo  {journal}
  {Phys. Rev. Lett.}\ }\textbf {\bibinfo {volume} {113}},\ \bibinfo {pages}
  {156402} (\bibinfo {year} {2014})}\BibitemShut {NoStop}%
\bibitem [{\citenamefont {Fr\'erot}\ \emph {et~al.}(2017)\citenamefont
  {Fr\'erot}, \citenamefont {Naldesi},\ and\ \citenamefont
  {Roscilde}}]{theory_entanglement3}%
  \BibitemOpen
  \bibfield  {author} {\bibinfo {author} {\bibfnamefont {I.}~\bibnamefont
  {Fr\'erot}}, \bibinfo {author} {\bibfnamefont {P.}~\bibnamefont {Naldesi}}, \
  and\ \bibinfo {author} {\bibfnamefont {T.}~\bibnamefont {Roscilde}},\ }\href
  {\doibase 10.1103/PhysRevB.95.245111} {\bibfield  {journal} {\bibinfo
  {journal} {Phys. Rev. B}\ }\textbf {\bibinfo {volume} {95}},\ \bibinfo
  {pages} {245111} (\bibinfo {year} {2017})}\BibitemShut {NoStop}%
\bibitem [{\citenamefont {Schachenmayer}\ \emph {et~al.}(2013)\citenamefont
  {Schachenmayer}, \citenamefont {Lanyon}, \citenamefont {Roos},\ and\
  \citenamefont {Daley}}]{theory_entanglement4}%
  \BibitemOpen
  \bibfield  {author} {\bibinfo {author} {\bibfnamefont {J.}~\bibnamefont
  {Schachenmayer}}, \bibinfo {author} {\bibfnamefont {B.~P.}\ \bibnamefont
  {Lanyon}}, \bibinfo {author} {\bibfnamefont {C.~F.}\ \bibnamefont {Roos}}, \
  and\ \bibinfo {author} {\bibfnamefont {A.~J.}\ \bibnamefont {Daley}},\ }\href
  {\doibase 10.1103/PhysRevX.3.031015} {\bibfield  {journal} {\bibinfo
  {journal} {Phys. Rev. X}\ }\textbf {\bibinfo {volume} {3}},\ \bibinfo {pages}
  {031015} (\bibinfo {year} {2013})}\BibitemShut {NoStop}%
\bibitem [{\citenamefont {Buyskikh}\ \emph {et~al.}(2016)\citenamefont
  {Buyskikh}, \citenamefont {Fagotti}, \citenamefont {Schachenmayer},
  \citenamefont {Essler},\ and\ \citenamefont {Daley}}]{theory_entanglement5}%
  \BibitemOpen
  \bibfield  {author} {\bibinfo {author} {\bibfnamefont {A.~S.}\ \bibnamefont
  {Buyskikh}}, \bibinfo {author} {\bibfnamefont {M.}~\bibnamefont {Fagotti}},
  \bibinfo {author} {\bibfnamefont {J.}~\bibnamefont {Schachenmayer}}, \bibinfo
  {author} {\bibfnamefont {F.}~\bibnamefont {Essler}}, \ and\ \bibinfo {author}
  {\bibfnamefont {A.~J.}\ \bibnamefont {Daley}},\ }\href {\doibase
  10.1103/PhysRevA.93.053620} {\bibfield  {journal} {\bibinfo  {journal} {Phys.
  Rev. A}\ }\textbf {\bibinfo {volume} {93}},\ \bibinfo {pages} {053620}
  (\bibinfo {year} {2016})}\BibitemShut {NoStop}%
\bibitem [{\citenamefont {Roy}\ and\ \citenamefont
  {Sharma}(2018)}]{theory_entanglement6}%
  \BibitemOpen
  \bibfield  {author} {\bibinfo {author} {\bibfnamefont {N.}~\bibnamefont
  {Roy}}\ and\ \bibinfo {author} {\bibfnamefont {A.}~\bibnamefont {Sharma}},\
  }\href {\doibase 10.1103/PhysRevB.97.125116} {\bibfield  {journal} {\bibinfo
  {journal} {Phys. Rev. B}\ }\textbf {\bibinfo {volume} {97}},\ \bibinfo
  {pages} {125116} (\bibinfo {year} {2018})}\BibitemShut {NoStop}%
\bibitem [{\citenamefont {Gong}\ \emph {et~al.}(2017)\citenamefont {Gong},
  \citenamefont {Foss-Feig}, \citenamefont {Brand\~ao},\ and\ \citenamefont
  {Gorshkov}}]{theory_entanglement7}%
  \BibitemOpen
  \bibfield  {author} {\bibinfo {author} {\bibfnamefont {Z.-X.}\ \bibnamefont
  {Gong}}, \bibinfo {author} {\bibfnamefont {M.}~\bibnamefont {Foss-Feig}},
  \bibinfo {author} {\bibfnamefont {F.~G. S.~L.}\ \bibnamefont {Brand\~ao}}, \
  and\ \bibinfo {author} {\bibfnamefont {A.~V.}\ \bibnamefont {Gorshkov}},\
  }\href {\doibase 10.1103/PhysRevLett.119.050501} {\bibfield  {journal}
  {\bibinfo  {journal} {Phys. Rev. Lett.}\ }\textbf {\bibinfo {volume} {119}},\
  \bibinfo {pages} {050501} (\bibinfo {year} {2017})}\BibitemShut {NoStop}%
\bibitem [{\citenamefont {Levitov}(1989)}]{theory_old_localization1}%
  \BibitemOpen
  \bibfield  {author} {\bibinfo {author} {\bibfnamefont {L.~S.}\ \bibnamefont
  {Levitov}},\ }\href {\doibase 10.1209/0295-5075/9/1/015} {\bibfield
  {journal} {\bibinfo  {journal} {Europhysics Letters ({EPL})}\ }\textbf
  {\bibinfo {volume} {9}},\ \bibinfo {pages} {83} (\bibinfo {year}
  {1989})}\BibitemShut {NoStop}%
\bibitem [{\citenamefont {Levitov}(1990)}]{theory_old_localization2}%
  \BibitemOpen
  \bibfield  {author} {\bibinfo {author} {\bibfnamefont {L.~S.}\ \bibnamefont
  {Levitov}},\ }\href {\doibase 10.1103/PhysRevLett.64.547} {\bibfield
  {journal} {\bibinfo  {journal} {Phys. Rev. Lett.}\ }\textbf {\bibinfo
  {volume} {64}},\ \bibinfo {pages} {547} (\bibinfo {year} {1990})}\BibitemShut
  {NoStop}%
\bibitem [{\citenamefont {Mirlin}\ \emph {et~al.}(1996)\citenamefont {Mirlin},
  \citenamefont {Fyodorov}, \citenamefont {Dittes}, \citenamefont {Quezada},\
  and\ \citenamefont {Seligman}}]{theory_old_localization3}%
  \BibitemOpen
  \bibfield  {author} {\bibinfo {author} {\bibfnamefont {A.~D.}\ \bibnamefont
  {Mirlin}}, \bibinfo {author} {\bibfnamefont {Y.~V.}\ \bibnamefont
  {Fyodorov}}, \bibinfo {author} {\bibfnamefont {F.-M.}\ \bibnamefont
  {Dittes}}, \bibinfo {author} {\bibfnamefont {J.}~\bibnamefont {Quezada}}, \
  and\ \bibinfo {author} {\bibfnamefont {T.~H.}\ \bibnamefont {Seligman}},\
  }\href {\doibase 10.1103/PhysRevE.54.3221} {\bibfield  {journal} {\bibinfo
  {journal} {Phys. Rev. E}\ }\textbf {\bibinfo {volume} {54}},\ \bibinfo
  {pages} {3221} (\bibinfo {year} {1996})}\BibitemShut {NoStop}%
\bibitem [{\citenamefont {Burin}\ and\ \citenamefont
  {Maksimov}(1989)}]{theory_old_localization4}%
  \BibitemOpen
  \bibfield  {author} {\bibinfo {author} {\bibfnamefont {A.~L.}\ \bibnamefont
  {Burin}}\ and\ \bibinfo {author} {\bibfnamefont {L.~A.}\ \bibnamefont
  {Maksimov}},\ }\href
  {http://www.jetpletters.ac.ru/ps/1129/article_17116.shtml} {\bibfield
  {journal} {\bibinfo  {journal} {Sov. Phys. JETP Lett.}\ }\textbf {\bibinfo
  {volume} {50}},\ \bibinfo {pages} {338} (\bibinfo {year} {1989})}\BibitemShut
  {NoStop}%
\bibitem [{\citenamefont {de~Moura}\ \emph {et~al.}(2005)\citenamefont
  {de~Moura}, \citenamefont {Malyshev}, \citenamefont {Lyra}, \citenamefont
  {Malyshev},\ and\ \citenamefont
  {Dom\'{\i}nguez-Adame}}]{theory_old_localization5}%
  \BibitemOpen
  \bibfield  {author} {\bibinfo {author} {\bibfnamefont {F.~A. B.~F.}\
  \bibnamefont {de~Moura}}, \bibinfo {author} {\bibfnamefont {A.~V.}\
  \bibnamefont {Malyshev}}, \bibinfo {author} {\bibfnamefont {M.~L.}\
  \bibnamefont {Lyra}}, \bibinfo {author} {\bibfnamefont {V.~A.}\ \bibnamefont
  {Malyshev}}, \ and\ \bibinfo {author} {\bibfnamefont {F.}~\bibnamefont
  {Dom\'{\i}nguez-Adame}},\ }\href {\doibase 10.1103/PhysRevB.71.174203}
  {\bibfield  {journal} {\bibinfo  {journal} {Phys. Rev. B}\ }\textbf {\bibinfo
  {volume} {71}},\ \bibinfo {pages} {174203} (\bibinfo {year}
  {2005})}\BibitemShut {NoStop}%
\bibitem [{\citenamefont {Rodr{\'{\i}}guez}\ \emph {et~al.}(2000)\citenamefont
  {Rodr{\'{\i}}guez}, \citenamefont {Malyshev},\ and\ \citenamefont
  {Dom{\'{\i}}nguez-Adame}}]{theory_old_localization6}%
  \BibitemOpen
  \bibfield  {author} {\bibinfo {author} {\bibfnamefont {A.}~\bibnamefont
  {Rodr{\'{\i}}guez}}, \bibinfo {author} {\bibfnamefont {V.~A.}\ \bibnamefont
  {Malyshev}}, \ and\ \bibinfo {author} {\bibfnamefont {F.}~\bibnamefont
  {Dom{\'{\i}}nguez-Adame}},\ }\href {\doibase 10.1088/0305-4470/33/15/102}
  {\bibfield  {journal} {\bibinfo  {journal} {Journal of Physics A:
  Mathematical and General}\ }\textbf {\bibinfo {volume} {33}},\ \bibinfo
  {pages} {L161} (\bibinfo {year} {2000})}\BibitemShut {NoStop}%
\bibitem [{\citenamefont {Deng}\ \emph {et~al.}(2018)\citenamefont {Deng},
  \citenamefont {Kravtsov}, \citenamefont {Shlyapnikov},\ and\ \citenamefont
  {Santos}}]{theory_power_law_loc1}%
  \BibitemOpen
  \bibfield  {author} {\bibinfo {author} {\bibfnamefont {X.}~\bibnamefont
  {Deng}}, \bibinfo {author} {\bibfnamefont {V.~E.}\ \bibnamefont {Kravtsov}},
  \bibinfo {author} {\bibfnamefont {G.~V.}\ \bibnamefont {Shlyapnikov}}, \ and\
  \bibinfo {author} {\bibfnamefont {L.}~\bibnamefont {Santos}},\ }\href
  {\doibase 10.1103/PhysRevLett.120.110602} {\bibfield  {journal} {\bibinfo
  {journal} {Phys. Rev. Lett.}\ }\textbf {\bibinfo {volume} {120}},\ \bibinfo
  {pages} {110602} (\bibinfo {year} {2018})}\BibitemShut {NoStop}%
\bibitem [{\citenamefont {Celardo}\ \emph {et~al.}(2016)\citenamefont
  {Celardo}, \citenamefont {Kaiser},\ and\ \citenamefont
  {Borgonovi}}]{theory_power_law_loc2}%
  \BibitemOpen
  \bibfield  {author} {\bibinfo {author} {\bibfnamefont {G.~L.}\ \bibnamefont
  {Celardo}}, \bibinfo {author} {\bibfnamefont {R.}~\bibnamefont {Kaiser}}, \
  and\ \bibinfo {author} {\bibfnamefont {F.}~\bibnamefont {Borgonovi}},\ }\href
  {\doibase 10.1103/PhysRevB.94.144206} {\bibfield  {journal} {\bibinfo
  {journal} {Phys. Rev. B}\ }\textbf {\bibinfo {volume} {94}},\ \bibinfo
  {pages} {144206} (\bibinfo {year} {2016})}\BibitemShut {NoStop}%
\bibitem [{\citenamefont {Nosov}\ \emph {et~al.}(2019)\citenamefont {Nosov},
  \citenamefont {Khaymovich},\ and\ \citenamefont
  {Kravtsov}}]{theory_power_law_loc3}%
  \BibitemOpen
  \bibfield  {author} {\bibinfo {author} {\bibfnamefont {P.~A.}\ \bibnamefont
  {Nosov}}, \bibinfo {author} {\bibfnamefont {I.~M.}\ \bibnamefont
  {Khaymovich}}, \ and\ \bibinfo {author} {\bibfnamefont {V.~E.}\ \bibnamefont
  {Kravtsov}},\ }\href {\doibase 10.1103/PhysRevB.99.104203} {\bibfield
  {journal} {\bibinfo  {journal} {Phys. Rev. B}\ }\textbf {\bibinfo {volume}
  {99}},\ \bibinfo {pages} {104203} (\bibinfo {year} {2019})}\BibitemShut
  {NoStop}%
\bibitem [{\citenamefont {Gopalakrishnan}(2017)}]{theory_power_law_loc4}%
  \BibitemOpen
  \bibfield  {author} {\bibinfo {author} {\bibfnamefont {S.}~\bibnamefont
  {Gopalakrishnan}},\ }\href {\doibase 10.1103/PhysRevB.96.054202} {\bibfield
  {journal} {\bibinfo  {journal} {Phys. Rev. B}\ }\textbf {\bibinfo {volume}
  {96}},\ \bibinfo {pages} {054202} (\bibinfo {year} {2017})}\BibitemShut
  {NoStop}%
\bibitem [{\citenamefont {Deng}\ \emph {et~al.}(2019)\citenamefont {Deng},
  \citenamefont {Ray}, \citenamefont {Sinha}, \citenamefont {Shlyapnikov},\
  and\ \citenamefont {Santos}}]{theory_power_law_loc5}%
  \BibitemOpen
  \bibfield  {author} {\bibinfo {author} {\bibfnamefont {X.}~\bibnamefont
  {Deng}}, \bibinfo {author} {\bibfnamefont {S.}~\bibnamefont {Ray}}, \bibinfo
  {author} {\bibfnamefont {S.}~\bibnamefont {Sinha}}, \bibinfo {author}
  {\bibfnamefont {G.~V.}\ \bibnamefont {Shlyapnikov}}, \ and\ \bibinfo {author}
  {\bibfnamefont {L.}~\bibnamefont {Santos}},\ }\href {\doibase
  10.1103/PhysRevLett.123.025301} {\bibfield  {journal} {\bibinfo  {journal}
  {Phys. Rev. Lett.}\ }\textbf {\bibinfo {volume} {123}},\ \bibinfo {pages}
  {025301} (\bibinfo {year} {2019})}\BibitemShut {NoStop}%
\bibitem [{\citenamefont {Biddle}\ \emph {et~al.}(2011)\citenamefont {Biddle},
  \citenamefont {Priour}, \citenamefont {Wang},\ and\ \citenamefont
  {Das~Sarma}}]{theory_power_law_loc6}%
  \BibitemOpen
  \bibfield  {author} {\bibinfo {author} {\bibfnamefont {J.}~\bibnamefont
  {Biddle}}, \bibinfo {author} {\bibfnamefont {D.~J.}\ \bibnamefont {Priour}},
  \bibinfo {author} {\bibfnamefont {B.}~\bibnamefont {Wang}}, \ and\ \bibinfo
  {author} {\bibfnamefont {S.}~\bibnamefont {Das~Sarma}},\ }\href {\doibase
  10.1103/PhysRevB.83.075105} {\bibfield  {journal} {\bibinfo  {journal} {Phys.
  Rev. B}\ }\textbf {\bibinfo {volume} {83}},\ \bibinfo {pages} {075105}
  (\bibinfo {year} {2011})}\BibitemShut {NoStop}%
\bibitem [{\citenamefont {Balagurov}\ \emph {et~al.}(2004)\citenamefont
  {Balagurov}, \citenamefont {Malyshev},\ and\ \citenamefont
  {Dom\'{\i}nguez~Adame}}]{theory_power_law_loc7}%
  \BibitemOpen
  \bibfield  {author} {\bibinfo {author} {\bibfnamefont {D.~B.}\ \bibnamefont
  {Balagurov}}, \bibinfo {author} {\bibfnamefont {V.~A.}\ \bibnamefont
  {Malyshev}}, \ and\ \bibinfo {author} {\bibfnamefont {F.}~\bibnamefont
  {Dom\'{\i}nguez~Adame}},\ }\href {\doibase 10.1103/PhysRevB.69.104204}
  {\bibfield  {journal} {\bibinfo  {journal} {Phys. Rev. B}\ }\textbf {\bibinfo
  {volume} {69}},\ \bibinfo {pages} {104204} (\bibinfo {year}
  {2004})}\BibitemShut {NoStop}%
\bibitem [{\citenamefont {Malyshev}\ \emph {et~al.}(2004)\citenamefont
  {Malyshev}, \citenamefont {Malyshev},\ and\ \citenamefont
  {Dom\'{\i}nguez-Adame}}]{theory_power_law_loc8}%
  \BibitemOpen
  \bibfield  {author} {\bibinfo {author} {\bibfnamefont {A.~V.}\ \bibnamefont
  {Malyshev}}, \bibinfo {author} {\bibfnamefont {V.~A.}\ \bibnamefont
  {Malyshev}}, \ and\ \bibinfo {author} {\bibfnamefont {F.}~\bibnamefont
  {Dom\'{\i}nguez-Adame}},\ }\href {\doibase 10.1103/PhysRevB.70.172202}
  {\bibfield  {journal} {\bibinfo  {journal} {Phys. Rev. B}\ }\textbf {\bibinfo
  {volume} {70}},\ \bibinfo {pages} {172202} (\bibinfo {year}
  {2004})}\BibitemShut {NoStop}%
\bibitem [{\citenamefont {Xiong}\ and\ \citenamefont
  {Zhang}(2003)}]{theory_power_law_loc9}%
  \BibitemOpen
  \bibfield  {author} {\bibinfo {author} {\bibfnamefont {S.-J.}\ \bibnamefont
  {Xiong}}\ and\ \bibinfo {author} {\bibfnamefont {G.-P.}\ \bibnamefont
  {Zhang}},\ }\href {\doibase 10.1103/PhysRevB.68.174201} {\bibfield  {journal}
  {\bibinfo  {journal} {Phys. Rev. B}\ }\textbf {\bibinfo {volume} {68}},\
  \bibinfo {pages} {174201} (\bibinfo {year} {2003})}\BibitemShut {NoStop}%
\bibitem [{\citenamefont {de~Brito}\ \emph {et~al.}(2004)\citenamefont
  {de~Brito}, \citenamefont {Rodrigues},\ and\ \citenamefont
  {Nazareno}}]{theory_power_law_loc10}%
  \BibitemOpen
  \bibfield  {author} {\bibinfo {author} {\bibfnamefont {P.~E.}\ \bibnamefont
  {de~Brito}}, \bibinfo {author} {\bibfnamefont {E.~S.}\ \bibnamefont
  {Rodrigues}}, \ and\ \bibinfo {author} {\bibfnamefont {H.~N.}\ \bibnamefont
  {Nazareno}},\ }\href {\doibase 10.1103/PhysRevB.69.214204} {\bibfield
  {journal} {\bibinfo  {journal} {Phys. Rev. B}\ }\textbf {\bibinfo {volume}
  {69}},\ \bibinfo {pages} {214204} (\bibinfo {year} {2004})}\BibitemShut
  {NoStop}%
\bibitem [{\citenamefont {de~Moura}\ and\ \citenamefont
  {Lyra}(1998)}]{theory_power_law_loc12}%
  \BibitemOpen
  \bibfield  {author} {\bibinfo {author} {\bibfnamefont {F.~A. B.~F.}\
  \bibnamefont {de~Moura}}\ and\ \bibinfo {author} {\bibfnamefont {M.~L.}\
  \bibnamefont {Lyra}},\ }\href {\doibase 10.1103/PhysRevLett.81.3735}
  {\bibfield  {journal} {\bibinfo  {journal} {Phys. Rev. Lett.}\ }\textbf
  {\bibinfo {volume} {81}},\ \bibinfo {pages} {3735} (\bibinfo {year}
  {1998})}\BibitemShut {NoStop}%
\bibitem [{\citenamefont {Cao}\ \emph {et~al.}(2017)\citenamefont {Cao},
  \citenamefont {Rosso}, \citenamefont {Bouchaud},\ and\ \citenamefont
  {Le~Doussal}}]{theory_power_law_loc13}%
  \BibitemOpen
  \bibfield  {author} {\bibinfo {author} {\bibfnamefont {X.}~\bibnamefont
  {Cao}}, \bibinfo {author} {\bibfnamefont {A.}~\bibnamefont {Rosso}}, \bibinfo
  {author} {\bibfnamefont {J.-P.}\ \bibnamefont {Bouchaud}}, \ and\ \bibinfo
  {author} {\bibfnamefont {P.}~\bibnamefont {Le~Doussal}},\ }\href {\doibase
  10.1103/PhysRevE.95.062118} {\bibfield  {journal} {\bibinfo  {journal} {Phys.
  Rev. E}\ }\textbf {\bibinfo {volume} {95}},\ \bibinfo {pages} {062118}
  (\bibinfo {year} {2017})}\BibitemShut {NoStop}%
\bibitem [{\citenamefont {Yao}\ \emph {et~al.}(2014)\citenamefont {Yao},
  \citenamefont {Laumann}, \citenamefont {Gopalakrishnan}, \citenamefont
  {Knap}, \citenamefont {M\"uller}, \citenamefont {Demler},\ and\ \citenamefont
  {Lukin}}]{theory_mbl_long_range0}%
  \BibitemOpen
  \bibfield  {author} {\bibinfo {author} {\bibfnamefont {N.~Y.}\ \bibnamefont
  {Yao}}, \bibinfo {author} {\bibfnamefont {C.~R.}\ \bibnamefont {Laumann}},
  \bibinfo {author} {\bibfnamefont {S.}~\bibnamefont {Gopalakrishnan}},
  \bibinfo {author} {\bibfnamefont {M.}~\bibnamefont {Knap}}, \bibinfo {author}
  {\bibfnamefont {M.}~\bibnamefont {M\"uller}}, \bibinfo {author}
  {\bibfnamefont {E.~A.}\ \bibnamefont {Demler}}, \ and\ \bibinfo {author}
  {\bibfnamefont {M.~D.}\ \bibnamefont {Lukin}},\ }\href {\doibase
  10.1103/PhysRevLett.113.243002} {\bibfield  {journal} {\bibinfo  {journal}
  {Phys. Rev. Lett.}\ }\textbf {\bibinfo {volume} {113}},\ \bibinfo {pages}
  {243002} (\bibinfo {year} {2014})}\BibitemShut {NoStop}%
\bibitem [{\citenamefont {Burin}(2015{\natexlab{a}})}]{theory_mbl_long_range1}%
  \BibitemOpen
  \bibfield  {author} {\bibinfo {author} {\bibfnamefont {A.~L.}\ \bibnamefont
  {Burin}},\ }\href {\doibase 10.1103/PhysRevB.91.094202} {\bibfield  {journal}
  {\bibinfo  {journal} {Phys. Rev. B}\ }\textbf {\bibinfo {volume} {91}},\
  \bibinfo {pages} {094202} (\bibinfo {year} {2015}{\natexlab{a}})}\BibitemShut
  {NoStop}%
\bibitem [{\citenamefont {Burin}(2015{\natexlab{b}})}]{theory_mbl_long_range2}%
  \BibitemOpen
  \bibfield  {author} {\bibinfo {author} {\bibfnamefont {A.~L.}\ \bibnamefont
  {Burin}},\ }\href {\doibase 10.1103/PhysRevB.92.104428} {\bibfield  {journal}
  {\bibinfo  {journal} {Phys. Rev. B}\ }\textbf {\bibinfo {volume} {92}},\
  \bibinfo {pages} {104428} (\bibinfo {year} {2015}{\natexlab{b}})}\BibitemShut
  {NoStop}%
\bibitem [{\citenamefont {Tikhonov}\ and\ \citenamefont
  {Mirlin}(2018)}]{theory_mbl_long_range3}%
  \BibitemOpen
  \bibfield  {author} {\bibinfo {author} {\bibfnamefont {K.~S.}\ \bibnamefont
  {Tikhonov}}\ and\ \bibinfo {author} {\bibfnamefont {A.~D.}\ \bibnamefont
  {Mirlin}},\ }\href {\doibase 10.1103/PhysRevB.97.214205} {\bibfield
  {journal} {\bibinfo  {journal} {Phys. Rev. B}\ }\textbf {\bibinfo {volume}
  {97}},\ \bibinfo {pages} {214205} (\bibinfo {year} {2018})}\BibitemShut
  {NoStop}%
\bibitem [{\citenamefont {De~Tomasi}(2019)}]{theory_long_range_mbl5}%
  \BibitemOpen
  \bibfield  {author} {\bibinfo {author} {\bibfnamefont {G.}~\bibnamefont
  {De~Tomasi}},\ }\href {\doibase 10.1103/PhysRevB.99.054204} {\bibfield
  {journal} {\bibinfo  {journal} {Phys. Rev. B}\ }\textbf {\bibinfo {volume}
  {99}},\ \bibinfo {pages} {054204} (\bibinfo {year} {2019})}\BibitemShut
  {NoStop}%
\bibitem [{\citenamefont {Singh}\ \emph {et~al.}(2017)\citenamefont {Singh},
  \citenamefont {Moessner},\ and\ \citenamefont
  {Roy}}]{theory_long_range_mbl6}%
  \BibitemOpen
  \bibfield  {author} {\bibinfo {author} {\bibfnamefont {R.}~\bibnamefont
  {Singh}}, \bibinfo {author} {\bibfnamefont {R.}~\bibnamefont {Moessner}}, \
  and\ \bibinfo {author} {\bibfnamefont {D.}~\bibnamefont {Roy}},\ }\href
  {\doibase 10.1103/PhysRevB.95.094205} {\bibfield  {journal} {\bibinfo
  {journal} {Phys. Rev. B}\ }\textbf {\bibinfo {volume} {95}},\ \bibinfo
  {pages} {094205} (\bibinfo {year} {2017})}\BibitemShut {NoStop}%
\bibitem [{\citenamefont {Nag}\ and\ \citenamefont
  {Garg}(2019)}]{arti_garg_2019}%
  \BibitemOpen
  \bibfield  {author} {\bibinfo {author} {\bibfnamefont {S.}~\bibnamefont
  {Nag}}\ and\ \bibinfo {author} {\bibfnamefont {A.}~\bibnamefont {Garg}},\
  }\href {\doibase 10.1103/PhysRevB.99.224203} {\bibfield  {journal} {\bibinfo
  {journal} {Phys. Rev. B}\ }\textbf {\bibinfo {volume} {99}},\ \bibinfo
  {pages} {224203} (\bibinfo {year} {2019})}\BibitemShut {NoStop}%
\bibitem [{\citenamefont {Russomanno}\ \emph {et~al.}(2017)\citenamefont
  {Russomanno}, \citenamefont {Iemini}, \citenamefont {Dalmonte},\ and\
  \citenamefont {Fazio}}]{theory_fully_connected1}%
  \BibitemOpen
  \bibfield  {author} {\bibinfo {author} {\bibfnamefont {A.}~\bibnamefont
  {Russomanno}}, \bibinfo {author} {\bibfnamefont {F.}~\bibnamefont {Iemini}},
  \bibinfo {author} {\bibfnamefont {M.}~\bibnamefont {Dalmonte}}, \ and\
  \bibinfo {author} {\bibfnamefont {R.}~\bibnamefont {Fazio}},\ }\href
  {\doibase 10.1103/PhysRevB.95.214307} {\bibfield  {journal} {\bibinfo
  {journal} {Phys. Rev. B}\ }\textbf {\bibinfo {volume} {95}},\ \bibinfo
  {pages} {214307} (\bibinfo {year} {2017})}\BibitemShut {NoStop}%
\bibitem [{\citenamefont {Lerose}\ \emph
  {et~al.}(2018{\natexlab{b}})\citenamefont {Lerose}, \citenamefont {Marino},
  \citenamefont {\ifmmode \check{Z}\else
  \v{Z}\fi{}unkovi\ifmmode~\check{c}\else \v{c}\fi{}}, \citenamefont
  {Gambassi},\ and\ \citenamefont {Silva}}]{theory_fully_connected2}%
  \BibitemOpen
  \bibfield  {author} {\bibinfo {author} {\bibfnamefont {A.}~\bibnamefont
  {Lerose}}, \bibinfo {author} {\bibfnamefont {J.}~\bibnamefont {Marino}},
  \bibinfo {author} {\bibfnamefont {B.}~\bibnamefont {\ifmmode \check{Z}\else
  \v{Z}\fi{}unkovi\ifmmode~\check{c}\else \v{c}\fi{}}}, \bibinfo {author}
  {\bibfnamefont {A.}~\bibnamefont {Gambassi}}, \ and\ \bibinfo {author}
  {\bibfnamefont {A.}~\bibnamefont {Silva}},\ }\href {\doibase
  10.1103/PhysRevLett.120.130603} {\bibfield  {journal} {\bibinfo  {journal}
  {Phys. Rev. Lett.}\ }\textbf {\bibinfo {volume} {120}},\ \bibinfo {pages}
  {130603} (\bibinfo {year} {2018}{\natexlab{b}})}\BibitemShut {NoStop}%
\bibitem [{\citenamefont {Sciolla}\ and\ \citenamefont
  {Biroli}(2011)}]{theory_fully_connected3}%
  \BibitemOpen
  \bibfield  {author} {\bibinfo {author} {\bibfnamefont {B.}~\bibnamefont
  {Sciolla}}\ and\ \bibinfo {author} {\bibfnamefont {G.}~\bibnamefont
  {Biroli}},\ }\href {\doibase 10.1088/1742-5468/2011/11/p11003} {\bibfield
  {journal} {\bibinfo  {journal} {Journal of Statistical Mechanics: Theory and
  Experiment}\ }\textbf {\bibinfo {volume} {2011}},\ \bibinfo {pages} {P11003}
  (\bibinfo {year} {2011})}\BibitemShut {NoStop}%
\bibitem [{\citenamefont {Gutman}\ \emph {et~al.}(2016)\citenamefont {Gutman},
  \citenamefont {Protopopov}, \citenamefont {Burin}, \citenamefont {Gornyi},
  \citenamefont {Santos},\ and\ \citenamefont
  {Mirlin}}]{energy_transport_theory1}%
  \BibitemOpen
  \bibfield  {author} {\bibinfo {author} {\bibfnamefont {D.~B.}\ \bibnamefont
  {Gutman}}, \bibinfo {author} {\bibfnamefont {I.~V.}\ \bibnamefont
  {Protopopov}}, \bibinfo {author} {\bibfnamefont {A.~L.}\ \bibnamefont
  {Burin}}, \bibinfo {author} {\bibfnamefont {I.~V.}\ \bibnamefont {Gornyi}},
  \bibinfo {author} {\bibfnamefont {R.~A.}\ \bibnamefont {Santos}}, \ and\
  \bibinfo {author} {\bibfnamefont {A.~D.}\ \bibnamefont {Mirlin}},\ }\href
  {\doibase 10.1103/PhysRevB.93.245427} {\bibfield  {journal} {\bibinfo
  {journal} {Phys. Rev. B}\ }\textbf {\bibinfo {volume} {93}},\ \bibinfo
  {pages} {245427} (\bibinfo {year} {2016})}\BibitemShut {NoStop}%
\bibitem [{\citenamefont {Bluhm}\ \emph {et~al.}(2009)\citenamefont {Bluhm},
  \citenamefont {Koshnick}, \citenamefont {Bert}, \citenamefont {Huber},\ and\
  \citenamefont {Moler}}]{Persistent_experiment1}%
  \BibitemOpen
  \bibfield  {author} {\bibinfo {author} {\bibfnamefont {H.}~\bibnamefont
  {Bluhm}}, \bibinfo {author} {\bibfnamefont {N.~C.}\ \bibnamefont {Koshnick}},
  \bibinfo {author} {\bibfnamefont {J.~A.}\ \bibnamefont {Bert}}, \bibinfo
  {author} {\bibfnamefont {M.~E.}\ \bibnamefont {Huber}}, \ and\ \bibinfo
  {author} {\bibfnamefont {K.~A.}\ \bibnamefont {Moler}},\ }\href {\doibase
  10.1103/PhysRevLett.102.136802} {\bibfield  {journal} {\bibinfo  {journal}
  {Phys. Rev. Lett.}\ }\textbf {\bibinfo {volume} {102}},\ \bibinfo {pages}
  {136802} (\bibinfo {year} {2009})}\BibitemShut {NoStop}%
\bibitem [{\citenamefont {Castellanos-Beltran}\ \emph
  {et~al.}(2013)\citenamefont {Castellanos-Beltran}, \citenamefont {Ngo},
  \citenamefont {Shanks}, \citenamefont {Jayich},\ and\ \citenamefont
  {Harris}}]{Persistent_experiment2}%
  \BibitemOpen
  \bibfield  {author} {\bibinfo {author} {\bibfnamefont {M.~A.}\ \bibnamefont
  {Castellanos-Beltran}}, \bibinfo {author} {\bibfnamefont {D.~Q.}\
  \bibnamefont {Ngo}}, \bibinfo {author} {\bibfnamefont {W.~E.}\ \bibnamefont
  {Shanks}}, \bibinfo {author} {\bibfnamefont {A.~B.}\ \bibnamefont {Jayich}},
  \ and\ \bibinfo {author} {\bibfnamefont {J.~G.~E.}\ \bibnamefont {Harris}},\
  }\href {\doibase 10.1103/PhysRevLett.110.156801} {\bibfield  {journal}
  {\bibinfo  {journal} {Phys. Rev. Lett.}\ }\textbf {\bibinfo {volume} {110}},\
  \bibinfo {pages} {156801} (\bibinfo {year} {2013})}\BibitemShut {NoStop}%
\bibitem [{\citenamefont {L\'evy}\ \emph {et~al.}(1990)\citenamefont {L\'evy},
  \citenamefont {Dolan}, \citenamefont {Dunsmuir},\ and\ \citenamefont
  {Bouchiat}}]{Persistent_experiment3}%
  \BibitemOpen
  \bibfield  {author} {\bibinfo {author} {\bibfnamefont {L.~P.}\ \bibnamefont
  {L\'evy}}, \bibinfo {author} {\bibfnamefont {G.}~\bibnamefont {Dolan}},
  \bibinfo {author} {\bibfnamefont {J.}~\bibnamefont {Dunsmuir}}, \ and\
  \bibinfo {author} {\bibfnamefont {H.}~\bibnamefont {Bouchiat}},\ }\href
  {\doibase 10.1103/PhysRevLett.64.2074} {\bibfield  {journal} {\bibinfo
  {journal} {Phys. Rev. Lett.}\ }\textbf {\bibinfo {volume} {64}},\ \bibinfo
  {pages} {2074} (\bibinfo {year} {1990})}\BibitemShut {NoStop}%
\bibitem [{\citenamefont {Chandrasekhar}\ \emph {et~al.}(1991)\citenamefont
  {Chandrasekhar}, \citenamefont {Webb}, \citenamefont {Brady}, \citenamefont
  {Ketchen}, \citenamefont {Gallagher},\ and\ \citenamefont
  {Kleinsasser}}]{Persistent_experiment4}%
  \BibitemOpen
  \bibfield  {author} {\bibinfo {author} {\bibfnamefont {V.}~\bibnamefont
  {Chandrasekhar}}, \bibinfo {author} {\bibfnamefont {R.~A.}\ \bibnamefont
  {Webb}}, \bibinfo {author} {\bibfnamefont {M.~J.}\ \bibnamefont {Brady}},
  \bibinfo {author} {\bibfnamefont {M.~B.}\ \bibnamefont {Ketchen}}, \bibinfo
  {author} {\bibfnamefont {W.~J.}\ \bibnamefont {Gallagher}}, \ and\ \bibinfo
  {author} {\bibfnamefont {A.}~\bibnamefont {Kleinsasser}},\ }\href {\doibase
  10.1103/PhysRevLett.67.3578} {\bibfield  {journal} {\bibinfo  {journal}
  {Phys. Rev. Lett.}\ }\textbf {\bibinfo {volume} {67}},\ \bibinfo {pages}
  {3578} (\bibinfo {year} {1991})}\BibitemShut {NoStop}%
\bibitem [{\citenamefont {B{\"u}ttiker}\ \emph {et~al.}(1983)\citenamefont
  {B{\"u}ttiker}, \citenamefont {Imry},\ and\ \citenamefont
  {Landauer}}]{persistent1}%
  \BibitemOpen
  \bibfield  {author} {\bibinfo {author} {\bibfnamefont {M.}~\bibnamefont
  {B{\"u}ttiker}}, \bibinfo {author} {\bibfnamefont {Y.}~\bibnamefont {Imry}},
  \ and\ \bibinfo {author} {\bibfnamefont {R.}~\bibnamefont {Landauer}},\
  }\href {\doibase https://doi.org/10.1016/0375-9601(83)90011-7} {\bibfield
  {journal} {\bibinfo  {journal} {Physics Letters A}\ }\textbf {\bibinfo
  {volume} {96}},\ \bibinfo {pages} {365 } (\bibinfo {year}
  {1983})}\BibitemShut {NoStop}%
\bibitem [{\citenamefont {Cheung}\ \emph {et~al.}(1988)\citenamefont {Cheung},
  \citenamefont {Gefen}, \citenamefont {Riedel},\ and\ \citenamefont
  {Shih}}]{persistent2}%
  \BibitemOpen
  \bibfield  {author} {\bibinfo {author} {\bibfnamefont {H.-F.}\ \bibnamefont
  {Cheung}}, \bibinfo {author} {\bibfnamefont {Y.}~\bibnamefont {Gefen}},
  \bibinfo {author} {\bibfnamefont {E.~K.}\ \bibnamefont {Riedel}}, \ and\
  \bibinfo {author} {\bibfnamefont {W.-H.}\ \bibnamefont {Shih}},\ }\href
  {\doibase 10.1103/PhysRevB.37.6050} {\bibfield  {journal} {\bibinfo
  {journal} {Phys. Rev. B}\ }\textbf {\bibinfo {volume} {37}},\ \bibinfo
  {pages} {6050} (\bibinfo {year} {1988})}\BibitemShut {NoStop}%
\bibitem [{\citenamefont {Cheung}\ \emph {et~al.}(1989)\citenamefont {Cheung},
  \citenamefont {Riedel},\ and\ \citenamefont {Gefen}}]{persistent3}%
  \BibitemOpen
  \bibfield  {author} {\bibinfo {author} {\bibfnamefont {H.-F.}\ \bibnamefont
  {Cheung}}, \bibinfo {author} {\bibfnamefont {E.~K.}\ \bibnamefont {Riedel}},
  \ and\ \bibinfo {author} {\bibfnamefont {Y.}~\bibnamefont {Gefen}},\ }\href
  {\doibase 10.1103/PhysRevLett.62.587} {\bibfield  {journal} {\bibinfo
  {journal} {Phys. Rev. Lett.}\ }\textbf {\bibinfo {volume} {62}},\ \bibinfo
  {pages} {587} (\bibinfo {year} {1989})}\BibitemShut {NoStop}%
\bibitem [{\citenamefont {Montambaux}\ \emph {et~al.}(1990)\citenamefont
  {Montambaux}, \citenamefont {Bouchiat}, \citenamefont {Sigeti},\ and\
  \citenamefont {Friesner}}]{persistent4}%
  \BibitemOpen
  \bibfield  {author} {\bibinfo {author} {\bibfnamefont {G.}~\bibnamefont
  {Montambaux}}, \bibinfo {author} {\bibfnamefont {H.}~\bibnamefont
  {Bouchiat}}, \bibinfo {author} {\bibfnamefont {D.}~\bibnamefont {Sigeti}}, \
  and\ \bibinfo {author} {\bibfnamefont {R.}~\bibnamefont {Friesner}},\ }\href
  {\doibase 10.1103/PhysRevB.42.7647} {\bibfield  {journal} {\bibinfo
  {journal} {Phys. Rev. B}\ }\textbf {\bibinfo {volume} {42}},\ \bibinfo
  {pages} {7647} (\bibinfo {year} {1990})}\BibitemShut {NoStop}%
\bibitem [{\citenamefont {Bouzerar}\ \emph {et~al.}(1994)\citenamefont
  {Bouzerar}, \citenamefont {Poilblanc},\ and\ \citenamefont
  {Montambaux}}]{persistent5}%
  \BibitemOpen
  \bibfield  {author} {\bibinfo {author} {\bibfnamefont {G.}~\bibnamefont
  {Bouzerar}}, \bibinfo {author} {\bibfnamefont {D.}~\bibnamefont {Poilblanc}},
  \ and\ \bibinfo {author} {\bibfnamefont {G.}~\bibnamefont {Montambaux}},\
  }\href {\doibase 10.1103/PhysRevB.49.8258} {\bibfield  {journal} {\bibinfo
  {journal} {Phys. Rev. B}\ }\textbf {\bibinfo {volume} {49}},\ \bibinfo
  {pages} {8258} (\bibinfo {year} {1994})}\BibitemShut {NoStop}%
\bibitem [{\citenamefont {Kohn}(1964)}]{drude1}%
  \BibitemOpen
  \bibfield  {author} {\bibinfo {author} {\bibfnamefont {W.}~\bibnamefont
  {Kohn}},\ }\href {\doibase 10.1103/PhysRev.133.A171} {\bibfield  {journal}
  {\bibinfo  {journal} {Phys. Rev.}\ }\textbf {\bibinfo {volume} {133}},\
  \bibinfo {pages} {A171} (\bibinfo {year} {1964})}\BibitemShut {NoStop}%
\end{thebibliography}%

\end{document}